\documentclass[aps,prd,11pt,showpacs,nofootinbib,longbibliography,superscriptaddress,floatfix]{revtex4-2}

\usepackage[T1]{fontenc}
\usepackage[utf8]{inputenc}
\usepackage{lmodern}
\usepackage{amsmath,amssymb,amsfonts}
\usepackage{bm}
\usepackage{booktabs}  
\usepackage{array}     

\usepackage{mathtools}
\usepackage{siunitx}
\usepackage{graphicx}
\usepackage{xcolor}
\usepackage{microtype}
\usepackage{multirow}
\usepackage[strings]{underscore}
\usepackage[colorlinks=true,linkcolor=blue,citecolor=blue,urlcolor=blue]{hyperref}
\usepackage[nameinlink,capitalise]{cleveref}

\newcommand{\diff}{\mathrm{d}}
\newcommand{\ee}{\mathrm{e}}
\newcommand{\ii}{\mathrm{i}}

\newcommand{\ord}{\mathrm{ord}}
\newcommand{\ran}{\mathrm{ran}}
\newcommand{\Bord}{B_{\ord}}
\newcommand{\Bran}{B_{\ran}}
\newcommand{\Btot}{B_{\rm tot}}

\newcommand{\vect}[1]{\bm{#1}}

\newcommand{\todo}[1]{}

\begin{document}

\title{Polarized and unpolarized synchrotron emission from dark matter in extragalactic targets}

\author{Javier Reynoso-Cordova}
\affiliation{Istituto Nazionale di Fisica Nucleare, Sezione di Napoli, 
\\Complesso Universitario di Monte Sant'Angelo,
Via Cintia, 80126 Napoli, Italy}
\affiliation{Department of Physics, University of Alberta, CCIS 4-181, \\ Edmonton, Alberta T6G 2E1, Canada}

\author{Catherine Gibson}
\affiliation{Department of Physics and Santa Cruz Institute for Particle Physics (SCIPP), University of California, Santa Cruz, Santa Cruz, CA 95064, USA}

\author{Stefano Profumo}
\affiliation{Department of Physics and Santa Cruz Institute for Particle Physics (SCIPP), University of California, Santa Cruz, Santa Cruz, CA 95064, USA}

\date{\today}

\begin{abstract}
We compute 95\% confidence-level upper limits on the dark matter annihilation cross section and decay rate from both total-intensity and polarized synchrotron emission in five extragalactic targets: M31, the Large Magellanic Cloud (LMC), the Draco and Sculptor dwarf spheroidal galaxies, and the Coma cluster. Using Planck maps at 30, 44, and 70\,GHz, we solve the diffusion--loss equation for dark-matter-produced electrons and positrons numerically with \texttt{DRAGON} and integrate the resulting synchrotron emission along the line of sight with \texttt{HERMES}, computing both total-intensity and polarized-intensity maps for each target with target-specific magnetic-field, gas, and radiation-field environments. The 30\,GHz channel yields the most stringent constraints in all cases, and limits on annihilation or decay into $e^+e^-$ are stronger than those for $b\bar{b}$ due to the harder injected spectrum. For most targets the total-intensity and polarized limits are broadly comparable; the LMC is an exception, where Faraday depolarization in the turbulent disk suppresses the polarized signal relative to total intensity, making total intensity the primary estimator. Our results are robust against the choice of flux estimator and coordinate uncertainty. This work demonstrates that microwave polarimetry provides a complementary and largely independent probe of dark matter synchrotron emission in extragalactic targets.
\end{abstract}

\maketitle

\section{Introduction}\label{sec:intro}
 
Dark matter (DM) models in which the dark-sector particle couples to Standard
Model fermions generically predict a population of energetic electrons and
positrons produced by annihilation or decay.  Wherever those particles
propagate through a magnetized medium they radiate synchrotron emission,
turning any sufficiently dense and magnetized DM halo into a potential
radio source.  The prospect of detecting such emission has motivated an
extensive program of observations and theoretical work targeting a variety of
extragalactic environments over the past two decades.
 
The framework for predicting multiwavelength radiative signals from DM
annihilation products was developed early in the context of galaxy clusters,
in particular the Coma cluster, by Colafrancesco, Profumo \& Ullio
\cite{Colafrancesco2006}, who solved the full diffusion--loss equation for
$e^{\pm}$ transport and computed the resulting synchrotron, inverse-Compton,
and Sunyaev-Zel'dovich signals.  Their companion study applying the same
formalism to dwarf spheroidal (dSph) galaxies, with Draco as the primary
target, established dSphs as among the cleanest laboratories for DM radio
searches owing to their high mass-to-light ratios and negligible astrophysical
backgrounds \cite{Colafrancesco2007}.
 
Observational searches have since been carried out across a wide range of
targets and radio facilities.  Deep observations of Local Group dSphs with
the Australia Telescope Compact Array (ATCA) placed stringent limits on WIMP
annihilation and decay from six galaxies including Draco, Sculptor, Fornax,
and Carina \cite{Regis2015I,Regis2015II,Regis2015III}, while a dedicated
search in the Reticulum II dSph provided competitive bounds near the
thermal-relic cross section \cite{Regis2017RetII}.  Stacking of 23 dSphs
in TGSS-ADR1 data at 150~MHz yielded further limits at lower frequency
\cite{Basu2021}.  For the Large Magellanic Cloud (LMC), the deep ASKAP-EMU
image at 888~MHz delivered strong constraints that exclude the thermal-relic
cross section for WIMP masses below $\sim\,$few hundred GeV in favorable
magnetic-field scenarios \cite{Regis2021LMC}; subsequent multi-frequency
analyses have extended and cross-checked those results \cite{Chen2024}.
Galaxy clusters and groups have been studied by several groups as well; the
RX-DMFIT tool \cite{McDaniel2017} provides a standardized pipeline for
cluster and dSph analyses, and the non-detection of diffuse synchrotron from
clusters in planned LOFAR, ASKAP, and APERTIF surveys was shown to have the
potential to reach the thermal-relic cross section for masses below
$\sim\!100$~GeV \cite{Storm2017, Storm:2012ty}.
 
Alongside total-intensity searches, the \emph{polarized} synchrotron
component of DM-induced emission has received comparatively little attention,
despite offering a complementary and potentially more discriminating probe.
Synchrotron radiation from a power-law electron distribution is intrinsically
partially linearly polarized; the observable polarization fraction encodes the
degree of magnetic-field ordering and is reduced by Faraday depolarization
along the line of sight.  A DM-induced signal, which injects $e^{\pm}$ pairs
isotropically into the turbulent field of a magnetized halo, has a
characteristically low intrinsic polarization fraction, distinct from
emission produced by ordered large-scale fields.  The first use of
synchrotron polarization to constrain DM annihilation was carried out for the
\emph{Galactic} halo by Manconi, Cuoco \& Lesgourgues \cite{Manconi2022},
who showed with Planck polarization maps that the polarized signal can be
more constraining than the total intensity by roughly an order of magnitude
for leptophilic models.  No analogous study has been performed for
extragalactic targets.
 
The present work fills that gap.  We apply both total-intensity and
polarized synchrotron emission as independent probes of DM annihilation and
decay in five extragalactic targets: the Andromeda galaxy (M31), the Large
Magellanic Cloud (LMC), the classical dSphs Draco and Sculptor, and the Coma
cluster.  For each target we solve the steady-state diffusion--loss equation
numerically with \texttt{DRAGON} \cite{Evoli:2016xgn}, adopting
target-specific magnetic-field models, interstellar radiation fields, and gas
distributions.  Line-of-sight integration for both Stokes~$I$ and polarized
intensity $P = \sqrt{Q^2 + U^2}$ is performed with \texttt{HERMES}
\cite{Dundovic:2021ryb}, using a new integration module for polarized
synchrotron we have implemented following the scheme of \texttt{GALPROP}
\cite{Strong1998,Moskalenko1998,Porter2022} and \texttt{Hammurabi}
\cite{Waelkens2009,Steininger2019}.  We compare the predicted signals against
Planck maps at 30, 44, and 70~GHz and derive 95\% confidence-level upper
limits on $\langle\sigma v\rangle$ and the decay rate $\Gamma$ for
annihilation and decay into $b\bar{b}$ and $e^+e^-$.
 
Our analysis extends recent work on DM searches in the LMC
\cite{Regis2021LMC,Reynoso-Cordova:2025meg}, dSphs
\cite{Regis:2023rpm,Todarello:2023hdk}, and total-intensity Planck studies
\cite{Regis2021LMC,Chen2024} to the previously unexplored regime of microwave
polarimetry in extragalactic systems.  The LMC proves to be a special case:
strong internal Faraday depolarization in its turbulent disk suppresses the
polarized signal relative to the total intensity, so that the polarized maps
yield the most conservative constraints for that target rather than the most
stringent.  For all other targets the polarized and total-intensity limits are
broadly comparable, demonstrating that microwave polarimetry provides a
genuinely independent and complementary channel for DM indirect detection in
extragalactic environments.
 
The paper is organized as follows.  Sections~\ref{sec:basics}--\ref{sec:stokes}
review the synchrotron physics, DM source terms, electron transport, and
the Stokes-parameter radiative transfer.  Sections~\ref{sec:isrf_models}--\ref{sec:radio_limits_flux}
describe the astrophysical inputs for each target: interstellar radiation
fields, DM density profiles, magnetic-field models, gas distributions, and
the observed radio flux limits used in the analysis.
Section~\ref{sec:lmc_radio} reviews the observational phenomenology of
polarized and unpolarized diffuse emission in the LMC, which motivates the
treatment of that target.  The numerical methodology is described in
Section~\ref{sec:Numerical}, and the results and constraints are presented
in Section~\ref{sec:results}.  We conclude in Section~\ref{sec:conclusions}.

\section{Synchrotron basics: spectra, polarization, and depolarization}
\label{sec:basics}
 
\subsection{Single-particle emission and the power-law spectrum}
 
For a relativistic electron of Lorentz factor $\gamma$ gyrating in a
magnetic field $\vect{B}$, the characteristic (critical) synchrotron
frequency is \cite{Rybicki1979,Longair2011}
\begin{equation}
  \nu_c = \frac{3e}{4\pi m_e c}\,B_\perp\,\gamma^2,\qquad
  B_\perp \equiv |\vect{B}\times\hat{n}|,
  \label{eq:nuc}
\end{equation}
with the single-particle power spectrum peaking near $0.29\,\nu_c$.
For an isotropic pitch-angle distribution and a power-law electron
number density $N_e(E)=K\,E^{-p}$, integrating over the pitch-angle
distribution and summing over all electrons yields a volume emissivity
\begin{equation}
  j_\nu \propto K\,B_\perp^{(p+1)/2}\,\nu^{-(p-1)/2}.
  \label{eq:jnu}
\end{equation}
The spectral index of the emissivity is $\alpha = (p-1)/2$, so $j_\nu
\propto \nu^{-\alpha}$.  Equation~\eqref{eq:jnu} is central to
interpreting DM constraints: the synchrotron signal scales as
$B^{(p+1)/2}$, making the \emph{total} field strength (not merely the
ordered component) the relevant quantity for setting bounds on
$\langle\sigma v\rangle$ and $\Gamma$.  This distinction matters for the
LMC, where the ordered and total fields differ substantially
(Section~\ref{sec:lmc_radio}).
 
\subsection{Intrinsic linear polarization}
\label{sec:intrinsic_pol}
 
In a perfectly uniform magnetic field, synchrotron emission from a
power-law electron distribution is intrinsically linearly polarized at
the fraction \cite{Rybicki1979,Longair2011}
\begin{equation}
  \Pi_0 = \frac{p+1}{p+7/3},
  \label{eq:Pi0}
\end{equation}
which evaluates to $\Pi_0 \simeq 0.69$ for $p = 2$ and
$\Pi_0 \simeq 0.74$ for $p = 3$ \cite{Beck2013}.  These
values represent a theoretical ceiling; in practice the observable
polarization fraction is always lower, reduced by several competing
physical mechanisms that we enumerate below in order of increasing
frequency dependence.
 
\subsection{Geometric (wavelength-independent) depolarization}
\label{sec:geom_depol}
 
\paragraph{Field-disorder depolarization.}
When the magnetic field consists of an ordered (coherent)
component $\Bord$ and an isotropic turbulent component $\Bran$, the
net complex polarization from an ensemble of emitting cells partially
cancels.  For a beam that averages over many turbulent cells,
\cite{Sokoloff1998}
\begin{equation}
  \Pi = \Pi_0\,\frac{\Bord^2}{\Bord^2 + \Bran^2},
  \label{eq:Pi_disorder}
\end{equation}
with $\Btot^2 = \Bord^2 + \Bran^2$.  This reduction is independent of
observing wavelength and sets a hard ceiling on the polarization
fraction even at arbitrarily high frequencies.  For the targets studied
here the ordered fraction $\Bord/\Btot$ ranges from $\lesssim 0.3$
(LMC disk) to $\sim 0.5$--$0.7$ (M31 ring) to values approaching
unity in the tidal filaments and dwarf-halo outskirts.
 
\paragraph{Beam depolarization.}
At angular resolutions that do not resolve the turbulent correlation
length of the field, regions with different polarization angles are
averaged within the synthesized beam.  This produces wavelength-independent
suppression that can mimic field disorder \cite{Sokoloff1998,Beck2013}.
At Planck frequencies and angular resolution ($\sim 5'$--$10'$) the beam
subtends scales of several kpc in all five targets, so beam depolarization
from unresolved turbulence is always present.
 
\paragraph{Thermal dilution.}
Free-free (bremsstrahlung) emission carries no net linear polarization.
For a thermal fraction $f_\mathrm{th}$ at a given observing frequency,
the observable polarization fraction is reduced by the factor
$(1 - f_\mathrm{th})$ relative to the purely nonthermal value.  In the
LMC, $f_\mathrm{th} \approx 30\%$ at 1.4~GHz \cite{Hassani2022}; at
the Planck frequencies used here (30--70~GHz) the free-free contribution
is smaller but non-negligible in actively star-forming targets.
 
\subsection{Faraday rotation and wavelength-dependent depolarization}
\label{sec:faraday}
 
A polarized radio wave propagating through a magneto-ionic medium
accumulates a rotation of its polarization angle:
\begin{equation}
  \Delta\chi(\lambda) = \mathrm{RM}\,\lambda^2,\qquad
  \mathrm{RM} = 0.81\!\int\! n_e(\mathrm{cm}^{-3})\,
               B_\parallel(\mu\mathrm{G})\,\diff l(\mathrm{pc}),
  \label{eq:RM}
\end{equation}
where $n_e$ is the thermal electron density and $B_\parallel$ is the
line-of-sight magnetic field component.  When the rotation measure
varies—either across the telescope beam or along the line of sight—the
net complex polarization is suppressed.  Three distinct mechanisms
operate \cite{Burn1966,Sokoloff1998}:
 
\paragraph{Differential Faraday rotation (DFR).}
If the synchrotron-emitting region is also a Faraday-rotating medium
(i.e.\ the emitting and rotating plasma are co-spatial), radiation from
the far side of the layer suffers a larger rotation than radiation from
the near side.  For a uniform slab of line-of-sight depth $L$ with
constant regular field and thermal electron density, the resulting
depolarization factor is \cite{Burn1966}
\begin{equation}
  \mathcal{D}_\mathrm{DFR} = \frac{\sin(2\,\mathrm{RM}_{\rm tot}\,\lambda^2)}
                                   {2\,\mathrm{RM}_{\rm tot}\,\lambda^2},
  \label{eq:DFR}
\end{equation}
where $\mathrm{RM}_{\rm tot}$ is the total rotation measure of the
slab.  This becomes severe at $\lambda^2 \gtrsim 1/|\mathrm{RM}|$ and
is particularly important for galaxy disk targets at centimetre
wavelengths.  At the Planck frequencies of 30--70~GHz ($\lambda \approx
4$--$10$~mm), $|\mathrm{RM}|\,\lambda^2 \ll 1$ for all targets in our
sample, so DFR is negligible and the polarization angle is preserved.
 
\paragraph{Faraday dispersion.}
Turbulent magnetic fields produce a dispersion $\sigma_\mathrm{RM}$ in
the rotation measure along and across the beam.  The polarized intensity
is then attenuated by \cite{Burn1966,Sokoloff1998}
\begin{equation}
  \mathcal{D}_\mathrm{FD} = \exp(-2\,\sigma_\mathrm{RM}^2\,\lambda^4).
  \label{eq:FD}
\end{equation}
This depolarization is sensitive to the fourth power of wavelength and
is therefore decisive at centimetre and decimetre wavelengths (as in
the LMC disk at 20~cm), but is negligible at the millimetre wavelengths
of the Planck bands.
 
The combined effect of the above mechanisms on the Stokes $Q$ and $U$
emission from a magneto-ionic slab, including both an ordered and a
turbulent field component, is treated in full generality by
\citet{Sokoloff1998}; we adopt their formalism for the line-of-sight
integration in Section~\ref{sec:stokes}.
 
\paragraph{Summary for this work.}
At Planck frequencies (30, 44, 70~GHz), the dominant polarization
suppression mechanisms for our five targets are (i) field-disorder
depolarization (Eq.~\ref{eq:Pi_disorder}) and (ii) beam depolarization,
both wavelength-independent.  Faraday effects (DFR and Faraday
dispersion, Eqs.~\ref{eq:DFR}--\ref{eq:FD}) play a negligible direct
numerical role at these frequencies, though they are critical for
understanding the phenomenology of the LMC at lower radio frequencies
(Section~\ref{sec:lmc_radio}) and for interpreting why the LMC
polarized maps are intrinsically faint at centimetre wavelengths.
 
\section{DM source terms and injection spectra}
\label{sec:source}
 
For annihilation into a Standard Model final state $f$, the $e^\pm$
source term is
\begin{equation}
  Q_{\rm ann}(E,\vect{r}) = \frac{\langle\sigma v\rangle}{2\,m_\chi^2}\,
  \rho^2(\vect{r})\,\frac{\diff N_e}{\diff E}(E),
  \label{eq:Qann}
\end{equation}
while for decay with rate $\Gamma = 1/\tau$,
\begin{equation}
  Q_{\rm dec}(E,\vect{r}) = \frac{\Gamma}{m_\chi}\,
  \rho(\vect{r})\,\frac{\diff N_e}{\diff E}(E).
  \label{eq:Qdec}
\end{equation}
Here $\rho(\vect{r})$ is the dark matter density profile (NFW or
Burkert, as specified per target in Section~\ref{sec:dm_profiles}) and
$\diff N_e/\diff E$ is the $e^\pm$ energy spectrum per annihilation or
decay event.  We consider two benchmark channels: $\chi\chi\to b\bar{b}$
(hadronic, yielding a soft $e^\pm$ spectrum from $\pi^\pm$ decays in
the fragmentation cascade) and $\chi\chi\to e^+e^-$ (purely leptonic,
yielding a hard monochromatic-like spectrum).  In both cases, $\diff
N_e/\diff E$ is obtained from the PPPC4DMID tabulations of \citet{Cirelli2011},
which include electroweak corrections at high DM masses.  For decay,
the same spectra apply with $m_\chi \to m_\chi/2$ for each final-state
particle.  The $e^+e^-$ channel gives tighter constraints than $b\bar{b}$
across all targets because the harder spectrum produces proportionally more
synchrotron power at the Planck frequencies.
 
\section{Electron transport: diffusion and energy losses}
\label{sec:transport}
 
\subsection{Diffusion--loss equation}
 
The steady-state phase-space density $\psi(E,\vect{r})$ of DM-produced
$e^\pm$ satisfies \cite{Colafrancesco2006,Colafrancesco2007}
\begin{equation}
  -\nabla\!\cdot\!\big[D(E,\vect{r})\,\nabla\psi\big]
  + \frac{\partial}{\partial E}\big[b(E,\vect{r})\,\psi\big]
  = Q(E,\vect{r}),
  \label{eq:diffloss}
\end{equation}
where $D(E,\vect{r})$ is the spatial diffusion coefficient and $b(E,\vect{r})
= -\diff E/\diff t$ is the total energy-loss rate.
 
\subsection{Energy-loss rate}
 
The loss rate receives contributions from four processes:
\begin{equation}
  b(E,\vect{r}) = b_{\rm syn}+b_{\rm IC}+b_{\rm brem}+b_{\rm Coul},
\end{equation}
of which synchrotron and inverse Compton (IC) losses dominate at the
electron energies ($E \gtrsim$~few GeV) that are relevant for synchrotron
emission at Planck frequencies.  In the Thomson regime,
\begin{equation}
  b_{\rm rad}(E,\vect{r}) = \frac{4}{3}\,\sigma_T c\,\gamma^2
  \big(U_B(\vect{r})+U_{\rm rad}(\vect{r})\big),\qquad
  U_B \equiv \frac{\Btot^2}{8\pi},
  \label{eq:brad}
\end{equation}
where $U_{\rm rad}$ is the total radiation energy density (CMB plus
starlight/IR; see Section~\ref{sec:isrf_models}).  Bremsstrahlung and Coulomb losses become relevant only at
$E \lesssim$ few hundred MeV; at the electron energies relevant
for Planck-frequency synchrotron ($E \gtrsim$ few GeV) they are
sub-dominant for all five targets.
For the Coma cluster and Draco/Sculptor, where the magnetic energy density
is small ($B \lesssim 5\,\mu$G and $B \lesssim 0.3\,\mu$G respectively),
IC losses off the CMB ($U_\mathrm{CMB} \simeq 0.26$~eV~cm$^{-3}$)
dominate over synchrotron; for M31 and the LMC, $U_B$ and $U_\mathrm{CMB}$
are comparable or $U_B$ is larger in the inner regions.
 
\subsection{Calorimetric and diffusion-dominated limits}
 
Two limiting regimes are useful for physical intuition:
\begin{itemize}
  \item \emph{Calorimetric limit} (loss-dominated, $\lambda \ll R_{\rm halo}$):
    the electrons lose all their energy locally, $\psi \simeq Q/b$,
    and the synchrotron emissivity traces the DM density squared
    ($\propto \rho^2$ for annihilation).
  \item \emph{Diffusion-dominated limit} ($\lambda \gg R_{\rm halo}$):
    the electrons propagate over a scale
    \begin{equation}
      \lambda^2(E,E_s) = 4\int_E^{E_s}\!\frac{D(E')}{b(E')}\,\diff E',
      \label{eq:lambda}
    \end{equation}
    before losing their energy, smoothing the emissivity relative to $Q$.
\end{itemize}
Galaxy clusters (Coma) with their high magnetic and radiation energy
densities are close to the calorimetric limit.  Dwarf spheroidals
(Draco, Sculptor), which have very low $U_B$ and $U_{\rm rad}$, are
more diffusion-dominated; the Planck-band synchrotron signal from these
targets is correspondingly spread over a larger solid angle.
 
All numerical solutions to Eq.~\eqref{eq:diffloss} in this work are
obtained with \texttt{DRAGON} \cite{Evoli:2016xgn}; the spatial-dependent
diffusion coefficient implemented there and the target-specific
parameters are detailed in Section~\ref{sec:Numerical}.
 
\section{From electrons to Stokes maps}
\label{sec:stokes}
 
Given the solved $\psi(E,\vect{r})$ and the target's magnetic-field
configuration $\vect{B}(\vect{r})$, we compute specific synchrotron
emissivities for Stokes $I$, $Q$, and $U$.  For an isotropic pitch-angle
distribution, the total-intensity emissivity at frequency $\nu$ is
\cite{Rybicki1979,Longair2011}
\begin{equation}
  j_{I,\nu}(\vect{r}) = \int_0^\infty \psi(E,\vect{r})\,
                         P_{\rm syn}(\nu,E,\vect{r})\,\diff E,
  \label{eq:jI}
\end{equation}
where $P_{\rm syn}$ is the single-electron synchrotron power integrated
over pitch angles.  The Stokes $Q$ and $U$ emissivities follow from the
same kernel evaluated for the ordered field component only, weighted by
the intrinsic polarization fraction $\Pi_0$ and the local polarization
angle $\chi_0(\vect{r})$ of the projected magnetic field
\cite{Waelkens2009,Steininger2019}.
 
Observable sky maps are obtained by integrating along the line of sight
with radiative transfer in the magneto-ionic medium.  In the limit where
Faraday effects are small (as at Planck frequencies; see
Section~\ref{sec:faraday}), the observable intensities simplify to
\begin{align}
  I_\nu(\hat{n}) &= \int j_{I,\nu}(\vect{r})\,\diff s,
  \label{eq:losI}\\
  Q_\nu(\hat{n})+\ii\,U_\nu(\hat{n}) &=
    \int \Pi_0\,j_{I,\nu}(\vect{r})\,
    \ee^{2\ii\chi_0(\vect{r})}\,\diff s
    \times \Big\langle \ee^{2\ii\lambda^2\mathrm{RM}(\vect{r})}\Big\rangle_{\rm beam},
  \label{eq:losQU}
\end{align}
where the beam average in Eq.~\eqref{eq:losQU} encodes residual
depolarization from line-of-sight and beam RM fluctuations via
Eqs.~\eqref{eq:Pi_disorder} and \eqref{eq:FD}.  The observable
polarized intensity is
\begin{equation}
  P_\nu(\hat{n}) = \sqrt{Q_\nu^2(\hat{n}) + U_\nu^2(\hat{n})},
  \label{eq:polint}
\end{equation}
which is the quantity compared to Planck maps in the analysis.  We
note that neither $V$ (circular polarization) nor the individual
Stokes $Q$ and $U$ maps are used as separate constraints; the analysis
operates on $I$ and $P$ independently.
 
The line-of-sight integrations in Eqs.~\eqref{eq:losI}--\eqref{eq:losQU}
are performed with \texttt{HERMES} \cite{Dundovic:2021ryb}, using a new
module for polarized synchrotron emission that we have implemented
following the numerical scheme of \texttt{GALPROP}
\cite{Strong1998,Moskalenko1998,Porter2022}, itself based on the
\texttt{Hammurabi} code \cite{Waelkens2009,Steininger2019}.  For M31
and the LMC, whose disk planes are inclined relative to the line of
sight, both the magnetic field and astrophysical source distributions
are rotated to align with the observed galaxy orientation before
integration, following the procedure of \citet{Reynoso-Cordova:2025meg}.

\section{Interstellar radiation fields and Gas Densities}
\label{sec:isrf_models}
 
The total radiation energy density entering the IC loss rate
(Eq.~\ref{eq:brad}) is decomposed as
\begin{equation}
  u_{\rm rad}(r) = u_{\rm CMB} + u_{\rm SL/IR}(r),
  \label{eq:urad}
\end{equation}
where $u_{\rm CMB} = aT_{\rm CMB}^4 \simeq 0.260~{\rm eV\,cm^{-3}}$
(with $T_{\rm CMB}=2.7255$~K) is spatially uniform and cosmologically
fixed,\footnote{All targets are at $z\approx0$; the tiny evolution of
$u_{\rm CMB}$ is irrelevant.} and $u_{\rm SL/IR}(r)$ encodes the
spatially varying starlight and dust-reprocessed infrared field from the
target's stellar and interstellar content.  We set the extragalactic
background light $u_{\rm EBL}=0$ throughout; including $u_{\rm
EBL}\lesssim 10^{-2}$~eV~cm$^{-3}$ changes results at the subpercent
level for all targets.
 
\subsection{M31 (Andromeda)}
 
M31's radiation field is dominated by its bulge and exponential disk.
Spatially resolved Spitzer/Herschel dust SED analyses provide maps of
the starlight heating rate, peaking in the central kpc and declining
over the disk \cite{Draine2014,Viaene2017}, consistent with structural
decompositions of the bulge and disk light \cite{Courteau2011}.  We
adopt a spherically averaged two-component profile:
\begin{equation}
u_{\rm SL/IR}^{\rm M31}(r) =
u_{b,0}\Big[1+\Big(\frac{r}{r_b}\Big)\Big]^{-\alpha_b}
+ u_{d,0}\exp\!\Big(-\frac{r}{R_d}\Big),
\label{eq:m31_isrf}
\end{equation}
with fiducial parameters
\begin{equation}
u_{b,0}=20~{\rm eV\,cm^{-3}},\quad r_b=0.6~{\rm kpc},\quad
\alpha_b=1.6;\qquad
u_{d,0}=0.6~{\rm eV\,cm^{-3}},\quad R_d=5.3~{\rm kpc}.
\end{equation}
These values yield $u_{\rm SL/IR}\sim 4$–$5$~eV~cm$^{-3}$ averaged
over the inner kpc and $\ll 1$~eV~cm$^{-3}$ beyond a few kpc,
consistent with the dust-heating analyses of \citet{Draine2014} and the
bulge/disk light profiles of \citet{Courteau2011}.

The cold ISM of M31, which enters the bremsstrahlung and Coulomb
loss rates in Eq.~\eqref{eq:diffloss}, is dominated by the 10-kpc
molecular ring in H$_2$ (traced by CO) and a broader H\,\textsc{i}
disk.  We model the 3D hydrogen nuclei density as
\begin{align}
n_{\rm H}^{\rm M31}(r,z) &=
  n_{\rm HI,0}\,e^{-r/R_{\rm HI}}\,
  \mathrm{sech}^2\!\!\Big(\frac{z}{h_{\rm HI}}\Big)\,
  f_{\rm hole}(r) \nonumber\\
&\quad +\, 2\,n_{\rm H_2,0}\,
  \exp\!\Big[-\frac{(r-r_{\rm ring})^2}{2\sigma_{\rm ring}^2}\Big]\,
  \mathrm{sech}^2\!\!\Big(\frac{z}{h_{\rm H_2}}\Big),
\label{eq:m31_gas}
\end{align}
with $f_{\rm hole}(r)\equiv 1-\exp[-(r/r_{\rm hole})^m]$ allowing a
central H\,\textsc{i} depression and fiducial scale parameters
$R_{\rm HI}=7$~kpc, $h_{\rm HI}=0.35$~kpc, $r_{\rm hole}=5$~kpc,
$m=2$; $r_{\rm ring}=10$~kpc, $\sigma_{\rm ring}=1.5$~kpc,
$h_{\rm H_2}=0.10$~kpc, anchored to the CO and H\,\textsc{i}
surface-density maps of \citet{Nieten2006}, \citet{Braun2009}, and
\citet{Corbelli2010}.  At the Planck-relevant electron energies
($E \gtrsim$ few~GeV), bremsstrahlung and Coulomb losses are
sub-dominant relative to synchrotron and IC, so the gas model
affects results only weakly and primarily at the lowest DM masses.

\subsection{LMC}
 
For the LMC we adopt the spatially uniform ISRF model of
\citet{Reynoso-Cordova:2025meg}, decomposed as a sum of five Planck
distributions following \citet{Acharyya2023}:
\begin{equation}
  u(\nu) = \sum_{k=1}^{5} \frac{U_k}{aT_k^4}\,u_P(\nu,T_k),
  \label{eq:lmc_isrf}
\end{equation}
with temperatures $T_k = \{2.73, 35, 330, 3800, 35000\}$~K and
energy densities 
$U_k = \{0.26, 0.12, \\ 0.025, 0.30, 1.20\}$~eV~cm$^{-3}$,
where $a = 4\sigma/c$.  The five components represent the CMB, the
far-infrared dust emission, mid-infrared emission, near-infrared stellar
emission, and optical/UV radiation from young stars, respectively.
The total $u_{\rm rad}$ integrated over this model is
$\sum_k U_k \simeq 1.94$~eV~cm$^{-3}$, roughly seven times larger than
$u_{\rm CMB}$ alone, reflecting the LMC's active star formation.

The LMC gas distribution is treated within the same
\texttt{DRAGON} framework used for the ISRF and magnetic field,
following \citet{Reynoso-Cordova:2025meg}; bremsstrahlung and
Coulomb losses are negligible for the electron energies relevant
to Planck-frequency synchrotron emission.

\subsection{Coma cluster}
 
In galaxy clusters the CMB is both spatially uniform and large
($u_{\rm CMB} \simeq 0.26$~eV~cm$^{-3}$), so it dominates the IC
losses over nearly the entire cluster volume.  Reviews of cluster
nonthermal phenomena confirm that IC losses are CMB-dominated except
very close to the BCG \cite{BrunettiJones2014}.  We model the
starlight/ICL contribution with a $\beta$-like profile normalized so
that it equals $u_{\rm CMB}$ at the characteristic equipartition
radius $r_{\rm eq} \simeq 0.03\,r_{200}$:
\begin{equation}
u_{\rm SL/IR}^{\rm Coma}(r)=
u_c\Big[1+\Big(\frac{r}{r_c}\Big)^2\Big]^{-3\beta/2},\qquad
u_c = u_{\rm CMB}\,\Big[1+\Big(\frac{r_{\rm eq}}{r_c}\Big)^2\Big]^{3\beta/2},
\label{eq:coma_isrf}
\end{equation}
with $r_{200}\simeq 2$~Mpc, $r_{\rm eq} \approx 60$~kpc, $r_c=50$~kpc,
and $\beta=0.6$ \cite{Mirakhor2020,Uchida2016}.  Numerically
$u_c \approx 0.57$~eV~cm$^{-3}$, so $u_{\rm SL/IR} \gtrsim u_{\rm CMB}$
only within $r \lesssim 60$~kpc; at all larger radii we effectively have
$u_{\rm rad} \simeq u_{\rm CMB}$.

The ICM electron density entering both the magnetic-field scaling
(Eq.~\ref{eq:coma_B_scaling}) and the bremsstrahlung loss rate is
given by the same $\beta$-model (Eq.~\ref{eq:coma_ne_B}).  For the
energy-loss calculation we convert to total gas density assuming
primordial abundances ($X\simeq0.76$, $Y\simeq0.24$):
\begin{equation}
n_{\rm H}(r) \simeq 0.83\,n_e(r),\qquad
n_{\rm gas}(r) \equiv n_e + n_{\rm ions} \simeq 1.92\,n_e(r).
\label{eq:coma_conversions}
\end{equation}
IC losses off the CMB dominate over bremsstrahlung and synchrotron
losses throughout virtually the entire cluster volume, so the gas
model is relevant only within the dense core.

\subsection{Draco and Sculptor dSphs}
 
Classical dSphs have negligible internal starlight and are located well
away from the Milky Way disk, so the CMB is the dominant and essentially
sole IC target field throughout \cite{Colafrancesco2007,Regis2015II,Hu2024}.
We include a small Plummer-like core for completeness,
\begin{equation}
u_{\rm SL/IR}^{\rm dSph}(r)=u_{\ast,0}\Big[1+\Big(\frac{r}{r_\ast}\Big)^2\Big]^{-5/2},
\quad u_{\ast,0}=0.01~{\rm eV\,cm^{-3}},\ r_\ast=0.3~{\rm kpc},
\label{eq:dsph_isrf}
\end{equation}
but in practice $u_{\rm SL/IR}^{\rm dSph} \ll u_{\rm CMB}$ everywhere,
so $u_{\rm rad} \approx u_{\rm CMB}$ for both dSphs.

Classical dSphs are essentially devoid of cold gas; H\,\textsc{i}
searches report non-detections with $M_{\rm HI}\lesssim
\mathrm{few}\times10^3\,M_\odot$ \cite{Grcevich2009,Spekkens2014},
implying mean densities $\langle n_{\rm H}\rangle \ll
10^{-3}$~cm$^{-3}$ well below the level at which bremsstrahlung
or Coulomb losses are relevant.  We therefore set
$n_{\rm gas}^{\rm dSph} = 0$ throughout.

\section{Dark matter density profiles}
\label{sec:dm_profiles}
 
We model the spherically averaged DM density as an NFW profile
\cite{Navarro1997}:
\begin{equation}
\rho_{\rm NFW}(r) = \frac{\rho_s}{(r/r_s)(1+r/r_s)^2},
\label{eq:nfw}
\end{equation}
parameterized by the scale density $\rho_s$ and scale radius $r_s$.
For the standard $(M_{200}, c_{200})$ convention,
$M_{200} = \frac{4\pi}{3}\,200\,\rho_c\,r_{200}^3$,
$c_{200} \equiv r_{200}/r_s$, and
\begin{equation}
\rho_s = \frac{200}{3}\,\rho_c\,
\frac{c_{200}^3}{\ln(1+c_{200}) - c_{200}/(1+c_{200})},
\label{eq:nfw_norm}
\end{equation}
where $\rho_c$ is the $z\simeq0$ critical density.  NFW profiles
accurately describe a wide range of halo masses in $\Lambda$CDM
simulations \cite{Navarro1997,DuttonMaccio2014}.  For the dSphs, where
cusp-core degeneracy is significant, we additionally verify below that
our NFW choices are consistent with published kinematic constraints.
 
The actual numerical values of $(\rho_s, r_s)$ used in DRAGON for each
target are listed in Table~\ref{tab:dragon_parameters}.
 
\subsection{M31 (Andromeda)}
 
Rotation-curve and halo modeling for M31 favors NFW/Einasto halos with
$M_{200} \sim (0.8$–$1.1)\times10^{12}\,M_\odot$ and $r_{200}\sim
190$–$210$~kpc \cite{Tamm2012}.  We adopt an NFW profile with
\begin{equation}
M_{200}=1.0\times10^{12}\ M_\odot,\qquad c_{200}=10,\qquad
r_{200}\simeq 206~{\rm kpc},\ r_s\simeq 20.6~{\rm kpc},
\label{eq:m31_nfw}
\end{equation}
consistent with earlier NFW bulge+disk+halo fits
\cite{Geehan2006,Seigar2008} and recent mass reconstructions from
halo kinematics \cite{Tamm2012,Kafle2018,Zhang2024}.
 
\subsection{LMC}
 
We adopt the NFW profile derived by fitting the LMC's H\,\textsc{i}
rotation curve \cite{Regis2021LMC}: $r_s = 9.8$~kpc and
$\rho_s = 0.2$~GeV~cm$^{-3}$ (corresponding to $\rho_s \simeq
1.66\times10^7\,M_\odot\,{\rm kpc}^{-3}$), as listed in
Table~\ref{tab:dragon_parameters}.\footnote{The companion
paper \citet{Regis2021LMC} quotes $r_s = 9.16$~kpc; the value
$r_s=9.8$~kpc in Table~\ref{tab:dragon_parameters} reflects a minor
rounding and rescaling in the DRAGON implementation, well within the
observational uncertainty of the fit.}
 
\subsection{Coma cluster (A1656)}
 
Rich clusters are well described by NFW profiles with low concentration;
for Coma, X-ray, lensing, and dynamical analyses favor
$M_{200}\approx (1.2$–$1.5)\times10^{15}\,M_\odot$ and $c_{200}\simeq
3$–$5$ \cite{LokasMamon2003,Brilenkov2015,DuttonMaccio2014}.  We use
the fiducial parameters of \citet{Brilenkov2015},
\begin{equation}
M_{200}=1.29\times10^{15}\ M_\odot,\qquad c_{200}=3.9,
\end{equation}
giving $r_{200}\simeq 2.6$~Mpc and $r_s\simeq 0.67$~Mpc.  These are
consistent with kinematic mass-anisotropy modeling \cite{LokasMamon2003}
and the expected $c(M)$ relation from $\Lambda$CDM simulations
\cite{DuttonMaccio2014}.
 
\subsection{Draco and Sculptor dSphs}
\label{sec:dsph_profiles}
 
Kinematic modeling of classical dSphs is consistent with both cuspy
(NFW) and cored (Burkert) inner slopes within current uncertainties;
robust constraints are typically expressed in terms of enclosed masses
or $J$-factors \cite{Strigari2014,Chiappo2019}.  For this work we use
NFW profiles with $(\rho_s, r_s)$ taken from the median posteriors of
published MCMC Jeans analyses of the stellar kinematics.  For Draco we
adopt the parameters of \citet{Regis:2023rpm}, and for Sculptor those
of \citet{Todarello:2023hdk}; the resulting values are listed in
Table~\ref{tab:dragon_parameters}.  These choices are consistent with
GravSphere modeling of the Draco cusp \cite{Read2018} and with
two-population Jeans fits to Sculptor \cite{Strigari2014}, and
reproduce the observationally well-constrained enclosed mass
$M(\leq300~{\rm pc})\sim 10^7\,M_\odot$ for both systems.
 
\section{Magnetic field models}
\label{sec:Bfields}
 
The magnetic field amplitude $B(r)$ enters both the synchrotron
emissivity ($j_\nu \propto B^{(p+1)/2}$) and the radiative energy-loss
rate ($b_{\rm syn} \propto B^2$).  We adopt observation-anchored
parameterizations for each target.
 
\subsection{M31 (Andromeda)}
 
Multi-wavelength radio polarimetry and equipartition analyses of M31
reveal a ring-like emission torus with a nearly axisymmetric ordered
field peaking at $\sim\!8$–$12$~kpc.  Recent analyses by
\citet{Beck2025} using observations at 3.6, 6.2, and 20.5~cm
wavelengths yield average equipartition strengths in the torus of
$B_{\rm tot} = 6.3\pm0.2~\mu$G (total), $B_{\rm turb} = 5.4\pm0.2~\mu$G
(isotropic turbulent), and $B_{\rm ord} = 3.2\pm0.3~\mu$G (ordered in
the sky plane), with the field declining exponentially outward on a scale
length $R_B \simeq 14.5$~kpc \cite{Beck2025,Fletcher2004}.  We model the
spherically averaged total field as
\begin{equation}
B_{\rm tot}^{\rm M31}(r) = B_{\rm ref}\,
\exp\!\left(-\frac{r-r_{\rm ref}}{R_B}\right),\quad
B_{\rm ref}=6.3~\mu{\rm G},\ r_{\rm ref}=10~{\rm kpc},\
R_B=14.5~{\rm kpc}.
\label{eq:m31_B}
\end{equation}
The ordered and turbulent fractions both decline with approximately the
same scale length, so the polarization fraction predicted by
Eq.~\eqref{eq:Pi_disorder} ($\Pi \approx \Pi_0 B_{\rm ord}^2/B_{\rm tot}^2
\simeq 0.26\,\Pi_0$) is approximately constant across the ring and
consistent with multi-frequency polarimetric observations
\cite{Beck2025,Fletcher2004}.
 
\subsection{LMC}
 
We adopt the double-exponential model of \citet{Reynoso-Cordova:2025meg},
\begin{equation}
B(r,z) = B_0\exp\!\left(-\frac{r}{r_{\rm scale}}\right)
          \exp\!\left(-\frac{|z|}{z_h}\right),
\label{eq:lmc_B}
\end{equation}
with cylindrical coordinates $(r,z)$ in the LMC frame, a central field
$B_0 = 4.3~\mu$G anchored by Faraday rotation measurements of polarized
background sources \cite{Gaensler2005}, and scale lengths $r_{\rm scale}
= 5$~kpc and $z_h = 1.5$~kpc.  The field orientation is taken to be
purely toroidal and axisymmetric, $\vec{B}(r,z) = B(r,z)\,\hat{\varphi}$.
This $B_0$ value reflects the \emph{total} field strength inferred from
the nonthermal synchrotron emissivity via energy equipartition
($\langle B\rangle \approx 10.1~\mu$G from Hassani et al.
\cite{Hassani2022}; the lower $B_0$ adopted here is appropriate for the
central region on the rotation-curve scale).  As discussed in
Section~\ref{sec:lmc_radio}, the ordered component is substantially
weaker than $B_{\rm tot}$ in the LMC disk; the total field is the
appropriate quantity for computing DM synchrotron constraints
(Section~\ref{sec:basics}).
 
\subsection{Coma cluster (A1656)}
 
Faraday RM modeling with VLA observations at 3.6, 6, and 20~cm
constrains the Coma magnetic field to follow a radial scaling tied to
the ICM electron density \cite{Bonafede2010}:
\begin{equation}
B^{\rm Coma}(r) = B_0\left[\frac{n_e(r)}{n_0}\right]^{\eta},\qquad
(B_0,\eta) \approx (4.7~\mu{\rm G},\,0.5).
\label{eq:coma_B_scaling}
\end{equation}
For the ICM electron density we use the standard $\beta$-model fit to
Coma X-ray observations \cite{Briel1992,Mohr1999}:
\begin{equation}
n_e(r) = n_0\left[1+\left(\frac{r}{r_c}\right)^2\right]^{-3\beta/2},\qquad
(n_0,\,r_c,\,\beta) = (3.36\times10^{-3}~{\rm cm^{-3}},\,310~{\rm kpc},\,0.75),
\label{eq:coma_ne_B}
\end{equation}
yielding
\begin{equation}
B^{\rm Coma}(r) = 4.7~\mu{\rm G}
\left[1+\left(\frac{r}{310~{\rm kpc}}\right)^2\right]^{-0.56}.
\label{eq:coma_B_profile}
\end{equation}
This prescription is the standard RM-calibrated choice for Coma
\cite{Bonafede2010,GovoniFeretti2004} and has been widely adopted in
DM radio searches targeting galaxy clusters.
 
\subsection{Draco and Sculptor dSphs}
 
Classical dSphs are gas-poor, quiescent, and lack detectable radio halos;
no direct measurement of their internal magnetic fields is available.
Radio non-detections and stacking limits place only upper bounds of
$B \lesssim$ few $\mu$G \cite{Regis2015II,Regis2015III,Basu2021,Natarajan2015}.
DM synchrotron studies in these targets typically bracket results across
$B = 0.1$–$1~\mu$G and find them CMB-loss dominated regardless of the
specific value \cite{Colafrancesco2007,Regis2015II}.  We adopt
\begin{equation}
B^{\rm dSph}(r) = B_\ast\left[1+\left(\frac{r}{r_\ast}\right)^2\right]^{-m/2},
\qquad B_\ast=0.3~\mu{\rm G},\ r_\ast=0.5~{\rm kpc},\ m=0,
\label{eq:dsph_B}
\end{equation}
i.e., a spatially uniform field of $0.3~\mu$G.  We have verified that
varying $B_\ast$ in the range $0.1$–$1~\mu$G shifts the resulting limits
by less than a factor of two, consistent with the insensitivity found in
earlier analyses \cite{Regis2015II,Basu2021}.

\section{Observational data and signal estimation}
\label{sec:radio_limits_flux}
 
\subsection{Primary data: Planck LFI maps}
\label{sec:planck_data}
 
The constraints derived in Section~\ref{sec:results} are based
exclusively on the Planck 2018 Low Frequency Instrument (LFI) full-sky
maps at 30, 44, and 70~GHz \cite{PlanckLFI2018}, which provide
co-registered Stokes $I$, $Q$, and $U$ maps in thermodynamic
($\mu\text{K}_\text{CMB}$) units.  The key instrumental properties are
summarized in Table~\ref{tab:planck_maps}.  For each target we extract
the signal at the source position and within a 0.5~deg aperture as
described in Section~\ref{sec:results}; the per-pixel noise entering the
likelihood (Eq.~\ref{eq:chi_squared}) is read directly from the Planck
noise covariance maps. Noise maps are generated by estimating the error at each pixel as the variance of all pixels within an annular region of 0.5 degrees at an $N_{\mathrm{side}} = 1024$. We propagate the error for the polarization noise map from the Q and U polarization noise maps. The noise map estimation follows the methodology described in Manconi et al. \cite{Manconi2022}. 
 
\begin{table}[t]
\centering
\caption{Planck 2018 LFI map properties used in this analysis
\cite{PlanckLFI2018}.  FWHM values are effective beam sizes.
The map noise $\sigma_\text{map}$ is the local per-pixel rms
read from the Planck noise covariance maps at the position of each
target; representative values are given in Table~\ref{tab:planck_obs}.}
\label{tab:planck_maps}
\begin{tabular}{cccc}
\toprule
$\nu$ [GHz] & Effective beam FWHM & Map units & Sky coverage \\
\midrule
30  & $32.3'$ & $\mu\text{K}_\text{CMB}$ & Full sky \\
44  & $27.0'$ & $\mu\text{K}_\text{CMB}$ & Full sky \\
70  & $13.2'$ & $\mu\text{K}_\text{CMB}$ & Full sky \\
\bottomrule
\end{tabular}
\end{table}
 
The polarized intensity map is formed as
$P = \sqrt{Q^2 + U^2}$ pixel-by-pixel, and is corrected for the
positive noise bias using the standard relation
$P_\text{debiased} = \sqrt{P^2 - 2\sigma_P^2}$ whenever
$P > \sqrt{2}\,\sigma_P$, with $\sigma_P^2 = (\sigma_Q^2 + \sigma_U^2)/2$.  
Table~\ref{tab:planck_obs} lists the observed values ($S_\text{obs}$)
and uncertainties ($\sigma_\text{error}$) at the center of each target
as extracted from the maps; these are the quantities entering
Eq.~\eqref{eq:chi_squared} in both the average and maximum-pixel
estimators described in Section~\ref{sec:results}.
 
\begin{table*}[t]
\centering
\caption{Planck 2018 LFI observed signals and uncertainties at the
center of each target, extracted from the total-intensity (Stokes~$I$)
and polarized-intensity ($P$) maps.  For each target and frequency,
the first row gives the average-pixel value and uncertainty within the
0.5~deg aperture; the second row gives the single-pixel value at the
nominal source coordinate.  All values are in $\mu\text{K}_\text{CMB}$.}
\label{tab:planck_obs}
\begin{tabular}{llcccc}
\toprule
Target & $\nu$ [GHz] & $S_\text{obs}^I$ [$\mu$K] & $\sigma_I$ [$\mu$K]
       & $S_\text{obs}^P$ [$\mu$K] & $\sigma_P$ [$\mu$K] \\
\midrule
\multirow{6}{*}{M31}
  & 30 & 169.7 & 78.0  & 19.0 & 94.2 \\
  &    & 158.1 &  5.0  & 18.2 &  6.0 \\
  & 44 & 70.2  & 77.4  &  5.96 & 109  \\
  &    & 65.7  &  5.0  &  5.59 &  6.98 \\
  & 70 & 44.5  & 66.6  &  9.68 & 88.1 \\
  &    & 42.6  &  4.3  &  8.48 &  5.71 \\
\midrule
\multirow{6}{*}{LMC}
  & 30 & 705   & 342   & 14.1 & 30.2 \\
  &    & 1016  & 14.9  & 16.4 &  0.753 \\
  & 44 & 280   & 155   &  5.71 & 48.1 \\
  &    & 414   &  7.11 &  5.71 &  1.16 \\
  & 70 & 137   & 88.0  &  3.34 & 27.3 \\
  &    & 200   &  4.66 &  2.84 &  0.712 \\
\midrule
\multirow{6}{*}{Draco}
  & 30 & 38.5  & 38.3  &  4.10 & 53.1 \\
  &    & 39.0  &  1.81 &  4.06 &  2.50 \\
  & 44 & 28.6  & 43.1  &  5.60 & 58.8 \\
  &    & 27.9  &  2.02 &  5.15 &  2.81 \\
  & 70 & 25.3  & 40.3  &  5.19 & 55.8 \\
  &    & 24.5  &  1.82 &  3.76 &  2.52 \\
\midrule
\multirow{6}{*}{Sculptor}
  & 30 & 26.7  & 55.6  &  1.59 & 76.5 \\
  &    & 25.7  &  3.00 &  2.98 &  4.15 \\
  & 44 & 22.6  & 61.1  &  1.34 & 96.8 \\
  &    & 21.8  &  3.32 &  2.69 &  5.29 \\
  & 70 & 21.2  & 55.2  &  2.13 & 80.6 \\
  &    & 22.3  &  2.97 &  2.13 &  4.27 \\
\midrule
\multirow{6}{*}{Coma}
  & 30 & 10.2  & 59.6  &  3.30 & 79.0 \\
  &    & 11.7  &  3.82 &  3.78 &  5.08 \\
  & 44 &  2.93 & 64.6  &  6.79 & 102  \\
  &    &  5.11 &  4.16 &  5.80 &  6.40 \\
  & 70 &  5.46 & 58.0  &  0.785& 79.5 \\
  &    &  7.70 &  3.72 &  2.58 &  5.10 \\
\bottomrule
\end{tabular}
\end{table*}
 
\subsection{Ancillary radio data}
\label{sec:ancillary_radio}
 
The Planck maps alone do not resolve the structure of the targets at
centimetre wavelengths or constrain the astrophysical environment at
the source.  Table~\ref{tab:ancillary_radio} summarizes the ancillary
radio observations used to inform the astrophysical modeling in
Sections~\ref{sec:isrf_models}--\ref{sec:Bfields} and to provide
external validation context.  None of these datasets enters the
likelihood of Eq.~\eqref{eq:chi_squared} directly.
 
\begin{table*}[t]
\centering
\caption{Ancillary radio observations used to constrain the
astrophysical environment of each target.  These data inform the
magnetic-field, gas, and ISRF models but are not the basis of
the DM constraints; those derive from Planck LFI maps
(Table~\ref{tab:planck_maps}).  Upper limits are $3\sigma$
unless noted.}
\label{tab:ancillary_radio}
\begin{tabular}{lccccl}
\toprule
Object & $\nu$ [GHz] & Beam (FWHM) & $\sigma_I$ [mJy/beam]
       & $\sigma_P$ [mJy/beam] & Reference \\
\midrule
\multicolumn{6}{l}{\textit{M31 (Andromeda) — disk/ring detected in Stokes $I$}} \\
M31 (WSRT) & 0.35 & ${\sim}4'$ & (ring detected) & rms from RM-synthesis maps
    & \cite{Giessubel2013M31_350MHz} \\
\midrule
\multicolumn{6}{l}{\textit{Coma cluster (A1656) — radio halo detected in Stokes $I$}} \\
Coma (SRT)        & 6.6 & $2.9'$ & 0.33 & 0.25
    & \cite{Murgia2024_Coma_SRT66} \\
Coma (SRT+VLA)    & 1.4 & $1'$--$3'$ & (halo detected)
    & (local map rms) & \cite{Murgia2024_Coma_SRT66} \\
\midrule
\multicolumn{6}{l}{\textit{Dwarf spheroidals — no diffuse emission detected}} \\
Draco (ATCA)    & 1.1--3.1 & ${\sim}1'$ & $\lesssim 0.05$ & $\lesssim 0.05$
    & \cite{Regis2015II} \\
Sculptor (ATCA) & 1.1--3.1 & ${\sim}1'$ & $\lesssim 0.05$ & $\lesssim 0.05$
    & \cite{Regis2015II} \\
dSph stack (TGSS-ADR1) & 0.150 & $25''$ & ${\sim}3.5$ & (not polarimetric)
    & \cite{Basu2021} \\
\bottomrule
\end{tabular}
\end{table*}
 
\paragraph{M31.}
The Gießübel et al.\ \cite{Giessubel2013M31_350MHz} WSRT 350~MHz
polarimetric mosaic detects diffuse emission from the M31 disk/ring in
total intensity.  The RM-synthesis maps provide per-beam polarization
noise $\sigma_P$ that constrains the ordered-field geometry used in our
magnetic-field model (Section~\ref{sec:Bfields}).
 
\paragraph{Coma.}
The Sardinia Radio Telescope (SRT) observations at 6.6~GHz
\cite{Murgia2024_Coma_SRT66} report $\sigma_I \simeq 0.33$ and
$\sigma_P \simeq 0.25$~mJy~beam$^{-1}$ at $2.9'$ resolution.  The
Coma radio halo is detected in Stokes~$I$ at 1.4~GHz with combined
SRT+VLA data; the absence of a detected polarized counterpart at those
frequencies is expected from the high internal Faraday depth of the
cluster (Section~\ref{sec:faraday}) and is consistent with the limits
from Bonafede et al.\ \cite{Bonafede2010}.
 
\paragraph{Draco and Sculptor.}
ATCA mosaics in the 1.1--3.1~GHz band \cite{Regis2015II}
reach $\sigma \lesssim 0.05$~mJy~beam$^{-1}$ rms without detecting
diffuse emission, placing a $3\sigma$ upper limit of
$0.15$~mJy~beam$^{-1}$ on any extended component at ${\sim}1'$
resolution.  These non-detections, together with the TGSS-ADR1
stacking result \cite{Basu2021} at 150~MHz
($5\sigma$ limit ${\sim}17$~mJy~beam$^{-1}$), bracket the radio
synchrotron flux from both dSphs across two decades in frequency and
are consistent with the magnetic field and DM constraints presented
in Sections~\ref{sec:Bfields} and \ref{sec:results}.

\section{Synchrotron polarization in the LMC: astrophysical context}
\label{sec:lmc_radio}

The LMC requires separate treatment from the other four targets because
the relationship between its total-intensity and polarized synchrotron
emission is inverted relative to what one might naively expect, with
direct consequences for how DM constraints are derived.

The total synchrotron emissivity scales as $j_\nu \propto B_\mathrm{tot}^{(p+1)/2}$
and traces the full magnetic energy density, including the isotropic
turbulent component.  The polarized intensity, by contrast, is
sensitive only to the ordered field component \cite{Beck2015}, and is
further attenuated by Faraday depolarization in magneto-ionic media
\cite{Burn1966,Sokoloff1998}.  In the LMC disk, which hosts a dense,
star-forming ISM with a turbulence-dominated total field
($\langle B\rangle \approx 10.1\,\mu$G, of which the isotropic
turbulent component dominates \cite{Hassani2022}), both mechanisms act
strongly at centimetre wavelengths.  The key observational consequence,
established by Gaensler et al.\ \cite{Gaensler2005} is that the diffuse polarized emission is spatially
\emph{anti}-correlated with the bright total-intensity features: the
star-forming disk is an efficient Faraday depolarizer, leaving only the
near-side halo — where the field is more ordered and the Faraday depth
is lower — detectable in polarization.  The ordered component is
substantially weaker than $B_\mathrm{tot}$ across most of the disk, and
the measured polarization fraction is accordingly low.

DM annihilation injects $e^+/e^-$ pairs isotropically into this turbulent
field, producing synchrotron emission with no intrinsic ordered component.
Any resulting signal propagating through the LMC disk is subject to the
same Faraday depolarization that suppresses the disk's own synchrotron
emission in polarization.  Consequently, the polarized maps of the LMC
are intrinsically faint relative to total intensity at the Planck
frequencies, and the upper limits derived from the polarized maps are
more stringent than those from total intensity — the opposite situation
to the other four targets, where the two estimators are broadly
comparable.  This makes total intensity the primary estimator for the
LMC DM constraints, with the polarized maps providing the most
conservative bound.

A further consequence concerns the choice of magnetic-field value
entering the synchrotron emissivity.  Since $j_\nu \propto
B_\mathrm{tot}^{(p+1)/2}$, the appropriate field is the total
equipartition value inferred from the nonthermal total intensity, not
the ordered-field value from RM analysis.  Using the latter would
systematically underestimate the predicted signal and weaken the
derived limits.

\section{Numerical methodology}
\label{sec:Numerical}
 
\subsection{Transport: \texttt{DRAGON}}
 
The steady-state $e^\pm$ phase-space density $\psi(E,\vect{r})$
satisfying Eq.~\eqref{eq:diffloss} is computed with the
\texttt{DRAGON} cosmic-ray transport code \cite{Evoli:2016xgn}.
For external galaxies with well-defined disk planes (M31 and LMC),
a 3D Cartesian implementation is required: the magnetic field, gas,
and ISRF distributions described in Sections~\ref{sec:isrf_models},
\ref{sec:Bfields}, and~\ref{sec:dm_profiles} are rotated into the
frame of each galaxy's observed inclination and position angle
before being passed to \texttt{DRAGON}, following the procedure
developed for the LMC in \citet{Reynoso-Cordova:2025meg}.  The
adopted inclination and position angles are
$(\Theta_\mathrm{LMC}, i_\mathrm{LMC}) = (122^\circ, 34^\circ)$
\cite{vanderMarel:2001qm} and
$(\Theta_\mathrm{M31}, i_\mathrm{M31}) = (38^\circ, 78^\circ)$
\cite{Kent_1989, walterbos1987multi}.
For Draco, Sculptor, and Coma, which lack a well-defined disk
plane, a spherically symmetric geometry suffices.
 
To ensure convergence of the 3D solutions, we adopt a
spatially dependent diffusion coefficient,
\begin{equation}
D(r,z,p) = D_0\,(p/p_0)^{\alpha_D}
            \exp\!\big[r/r_{\rm diff}\big]\,
            \exp\!\big[z/z_{\rm diff}\big],
\label{eq:D_spatial}
\end{equation}
where $r_{\rm diff}$ and $z_{\rm diff}$ confine the diffusion to the
region occupied by the magnetic field.  For M31 and the LMC,
$(r,z)$ are cylindrical coordinates aligned with the galaxy plane;
for the three spherically symmetric targets, $r_{\rm diff}$ alone
sets the diffusion boundary.  The numerical grid extents and all
free parameters are listed in Table~\ref{tab:dragon_parameters}.
 
\subsection{Line-of-sight integration: \texttt{HERMES}}
 
Given $\psi(E,\vect{r})$ from \texttt{DRAGON}, the Stokes~$I$
and polarized-intensity maps are computed by integrating
Eqs.~\eqref{eq:losI}--\eqref{eq:losQU} along lines of sight
with the \texttt{HERMES} code \cite{Dundovic:2021ryb}.  For M31
and the LMC, the magnetic field configuration is rotated to match
the observed disk orientation before integration.  We have
implemented a new \texttt{HERMES} module for polarized synchrotron
emission following the numerical scheme of
\texttt{GALPROP} \cite{Strong1998,Moskalenko1998,Porter2022},
which in turn derives from \texttt{Hammurabi}
\cite{Waelkens2009,Steininger2019}.  This yields self-consistent
$I$ and $P = \sqrt{Q^2+U^2}$ maps for all five targets within
a single computational framework.
 
The NFW halo injection source is spherically symmetric and
therefore invariant under the disk-plane rotation, simplifying
the coordinate bookkeeping for M31 and the LMC.
 
\begin{table}[htbp]
\centering
\caption{\texttt{DRAGON} numerical parameters for each target.
The grid spans $(x,y,z)_\mathrm{min/max}$ in physical kpc;
\texttt{dim} is the number of grid points per axis; $r_\mathrm{diff}$
and $z_\mathrm{diff}$ define the diffusion-region boundary
(Eq.~\ref{eq:D_spatial}); $p_0$ and $E_\mathrm{min}$ are the
momentum and energy normalizations; $\rho_s$ and $r_s$ are the
NFW halo parameters used in the injection source term.}
\label{tab:dragon_parameters}
\begin{tabular}{@{}lccccc@{}}
\toprule
\textbf{Parameter} & \textbf{M31} & \textbf{LMC}
  & \textbf{Draco} & \textbf{Sculptor} & \textbf{Coma} \\
\midrule
$(x,y)_{\min/\max}$ [kpc] & $\pm20$ & $\pm10$ & $\pm3$ & $\pm3$ & $\pm10^3$ \\
$z_{\min/\max}$ [kpc]      & $\pm5$  & $\pm2$  & $\pm3$ & $\pm3$ & $\pm10^3$ \\
Grid points per axis & 50 & 60 & 40 & 40 & 50 \\
$r_{\rm diff}$ [kpc] & 15   & 5    & 2    & 2    & 500  \\
$z_{\rm diff}$ [kpc] & 2.5  & 1.5  & ---  & ---  & ---  \\
$p_0$ [GeV]          & 1    & 1    & 1    & 1    & 1    \\
$E_{\min}$ [GeV]     & 1    & 1    & 1    & 1    & 1    \\
$\rho_s$ [GeV\,cm$^{-3}$] & 0.232 & 0.200 & 1.38 & 0.77 & 0.017 \\
$r_s$ [kpc]          & 20   & 9.8  & 1.12 & 1.5  & 670  \\
\bottomrule
\end{tabular}
\end{table}

\section{Analysis and results}
\label{sec:results}
 
\subsection{Statistical method}
 
To derive upper limits on the dark matter annihilation cross section
$\langle\sigma v\rangle$ and decay rate $\Gamma$ we employ a
single-bin profile likelihood.  For a given DM mass $m_\chi$ and
channel, the \texttt{DRAGON}+\texttt{HERMES} pipeline yields a
predicted sky map $S_\mathrm{model}(m_\chi, l, b)$ in
$\mu\mathrm{K}_\mathrm{CMB}$, normalized to a reference value
$\langle\sigma v\rangle_\mathrm{ref}$ or $\Gamma_\mathrm{ref}$.
We parametrize the overall signal amplitude by a free scaling factor
$\eta$,
\begin{equation}
  S(\eta) = \eta \cdot S_\mathrm{model}(m_\chi),
  \label{eq:signal_model}
\end{equation}
and adopt a Gaussian likelihood,
\begin{equation}
\mathcal{L}(\eta) = e^{-\chi^2/2},\qquad
\chi^2 = \frac{\bigl(S_\mathrm{obs} - \eta\,S_\mathrm{model}(m_\chi)\bigr)^2}
              {\sigma_\mathrm{map}^2},
\label{eq:chi_squared}
\end{equation}
where $S_\mathrm{obs}$ is the observed Planck map value at the target
position and $\sigma_\mathrm{map}$ is the corresponding per-pixel
noise read from the Planck noise covariance maps
(Table~\ref{tab:planck_obs}).
 
A key step is the choice of $S_\mathrm{model}$ entering
Eq.~\eqref{eq:signal_model}.  Because the DM signal peaks at the halo
center, we evaluate the predicted map at its maximum over the
$0.5^\circ$ aperture,
$S_\mathrm{model}(m_\chi) = \max_{(l,b)\in\Omega_{0.5}}\!
[S_\mathrm{model}(m_\chi, l, b)]$,
rather than at a specific pixel.  This makes the normalization
conservative with respect to coordinate mismatches between the model
center and the map pixel grid: if the true halo peak falls between
pixels, sampling the maximum of the model (rather than the single pixel
at the nominal coordinate) prevents the denominator of
Eq.~\eqref{eq:chi_squared} from being artificially large.
 
We set a $95\%$~CL upper limit on $\eta$ by requiring
$\Delta\chi^2 = \chi^2(\eta_\mathrm{lim}) - \chi^2(\eta_\mathrm{b.f.})
< 3.84$, where $\eta_\mathrm{b.f.}$ minimizes Eq.~\eqref{eq:chi_squared}
analytically.  The physical limits follow as
\begin{equation}
  \langle\sigma v\rangle_\mathrm{lim} = \langle\sigma v\rangle_\mathrm{ref}
  \cdot \eta_\mathrm{lim}, \qquad
  \Gamma_\mathrm{lim} = \Gamma_\mathrm{ref} \cdot \eta_\mathrm{lim}.
\end{equation}
The analysis is run independently for total-intensity ($S_\mathrm{obs}
= I$, $\sigma_\mathrm{map} = \sigma_I$) and polarized-intensity
($S_\mathrm{obs} = P$, $\sigma_\mathrm{map} = \sigma_P$) maps,
yielding two sets of constraints for each target and frequency.
 
\subsection{Signal estimators and robustness}
 
The choice of $S_\mathrm{obs}$ warrants care because the finite Planck
pixel size ($\sim 3.8'$ at HEALPix $N_\mathrm{side}=512$) and residual
optical-to-radio astrometric offsets can shift the observed emission
peak relative to the nominal target coordinate.  We employ three
estimators for each target: (i)~the pixel value at the nominal source
coordinate; (ii)~the mean pixel value within a $0.5^\circ$ aperture
centered on the source; and (iii)~the maximum pixel value within the
same aperture.  The $0.5^\circ$ radius encompasses either the full
predicted DM emission profile or at least the NFW scale radius for all
five targets.
 
\begin{figure}[t]
  \centering
  \includegraphics[width=0.49\textwidth]{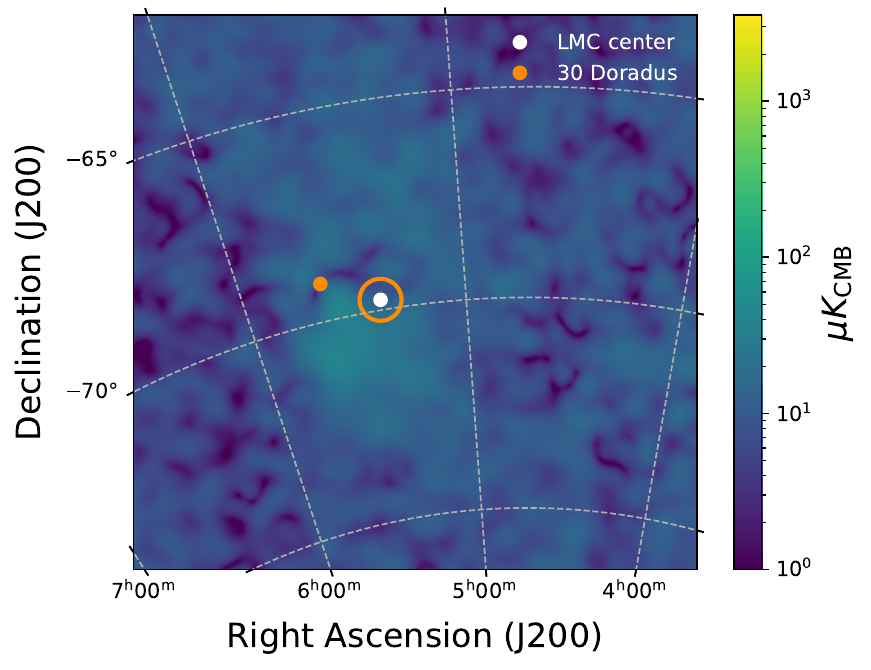}
  \hfill
  \includegraphics[width=0.49\textwidth]{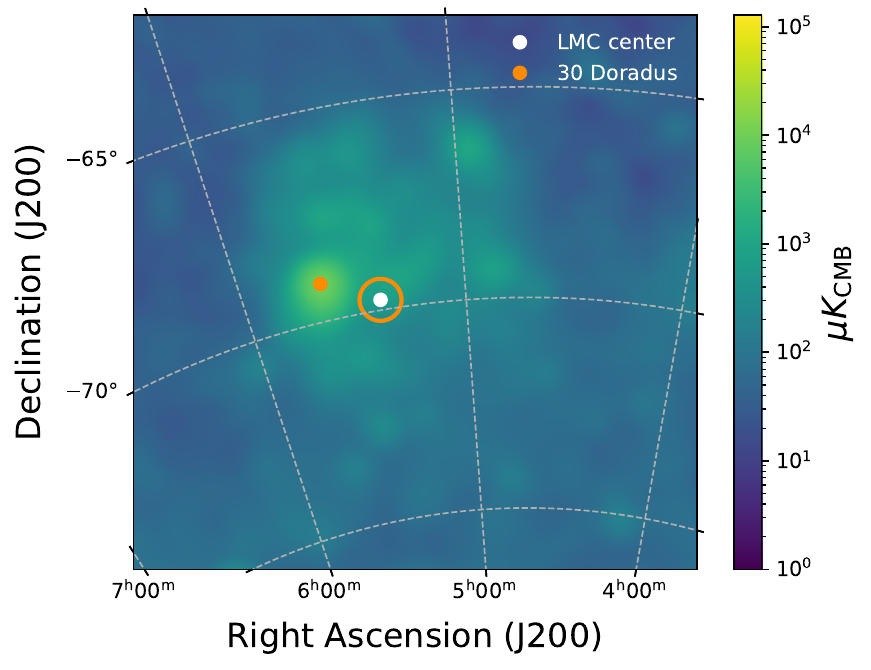}
  \caption{Planck 30~GHz maps centered on the LMC.
    \textit{Left:} Polarized intensity $P=\sqrt{Q^2+U^2}$.
    \textit{Right:} Total intensity $I$.
    The white dot marks the LMC optical center
    \cite{vanderMarel:2001qm} and the orange circle the $0.5^\circ$
    integration aperture.  The bright extended emission visible in
    total intensity to the upper left of the center is the 30~Doradus
    star-forming complex, which lies outside the aperture and does not
    contaminate the DM signal region.  The polarized map is markedly
    fainter than the total-intensity map across the disk, consistent
    with the Faraday depolarization discussed in
    Section~\ref{sec:lmc_radio}.}
  \label{fig:LMC_maps}
\end{figure}

The LMC requires special treatment.  At its distance of 49.9~kpc, the
$0.5^\circ$ aperture subtends only $\sim 0.44$~kpc, which is small
enough to exclude the bright 30~Doradus star-forming complex while
still capturing the dark matter emission near the halo center
\cite{Regis2021LMC}.  This is illustrated in
Figure~\ref{fig:LMC_maps}: the 30~Doradus region — the dominant
source of extended emission in total intensity — lies outside the
integration aperture (orange circle), whereas the optical center of the
LMC \cite{vanderMarel:2001qm} sits within it.
 
\begin{figure}[t]
  \centering
  \includegraphics[width=0.48\textwidth]{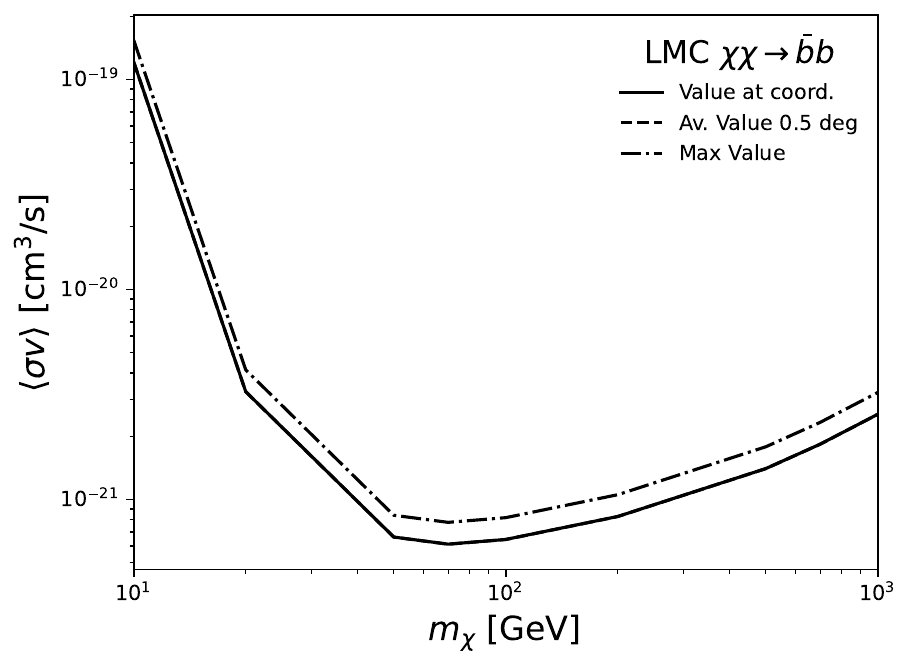}
  \hfill
  \includegraphics[width=0.48\textwidth]{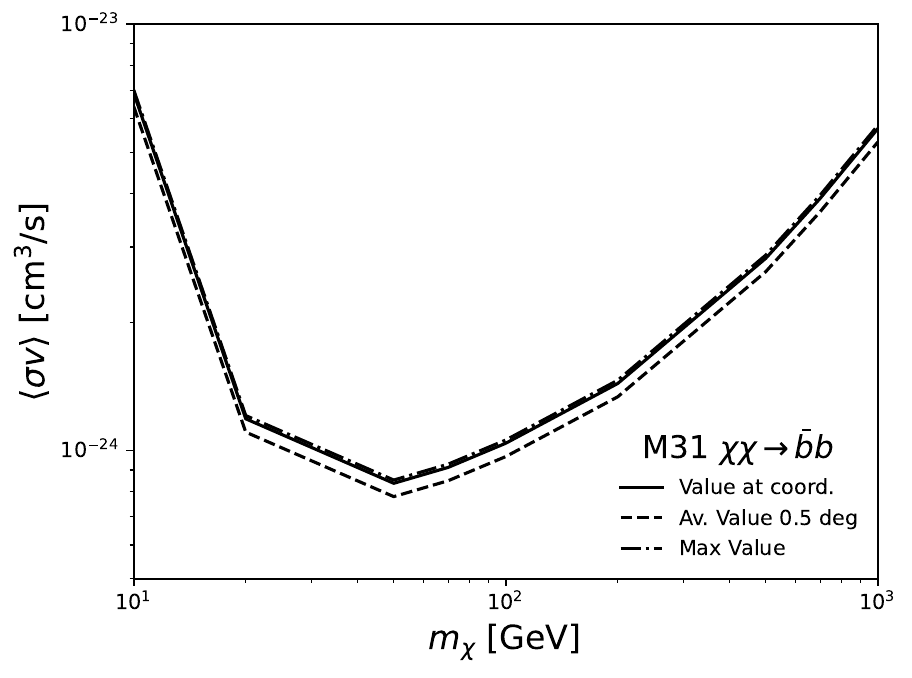}
  \caption{Comparison of the three flux estimators at 30~GHz for
    $b\bar{b}$ annihilation: the pixel value at the nominal source
    coordinate (solid), the mean pixel value within the $0.5^\circ$
    aperture (dashed), and the maximum pixel value within the same
    aperture (dash-dotted).  Left: LMC.  Right: M31.  The three
    estimators agree to within a factor of $\sim 2$ across the full
    mass range, with the maximum-pixel estimator producing the weakest
    constraints and the coordinate and aperture-average estimators
    agreeing closely.  The same level of consistency holds at 44 and
    70~GHz and for the $e^+e^-$ channel.}
  \label{fig:limits_robustness}
\end{figure}

Figure~\ref{fig:limits_robustness} compares the three estimators for
M31 and the LMC at 30~GHz for the $b\bar{b}$ channel.  Across the
full DM mass range, the three estimators agree to within a factor of
$\sim 2$.  As expected, the maximum-pixel estimator produces the
weakest constraints: a higher background flux requires a larger DM
signal to produce a statistically significant excess, relaxing the
upper limit.  The coordinate-pixel and aperture-average estimators
yield the most stringent limits and agree closely with each other.
This consistency demonstrates that the results are robust against
coordinate uncertainties and the specific choice of integration region.
We have verified that the same level of agreement holds at 44 and
70~GHz and for the $e^+e^-$ channel.
 
\subsection{Results}
 
\begin{figure}[t]
  \centering
  \includegraphics[width=0.32\textwidth]{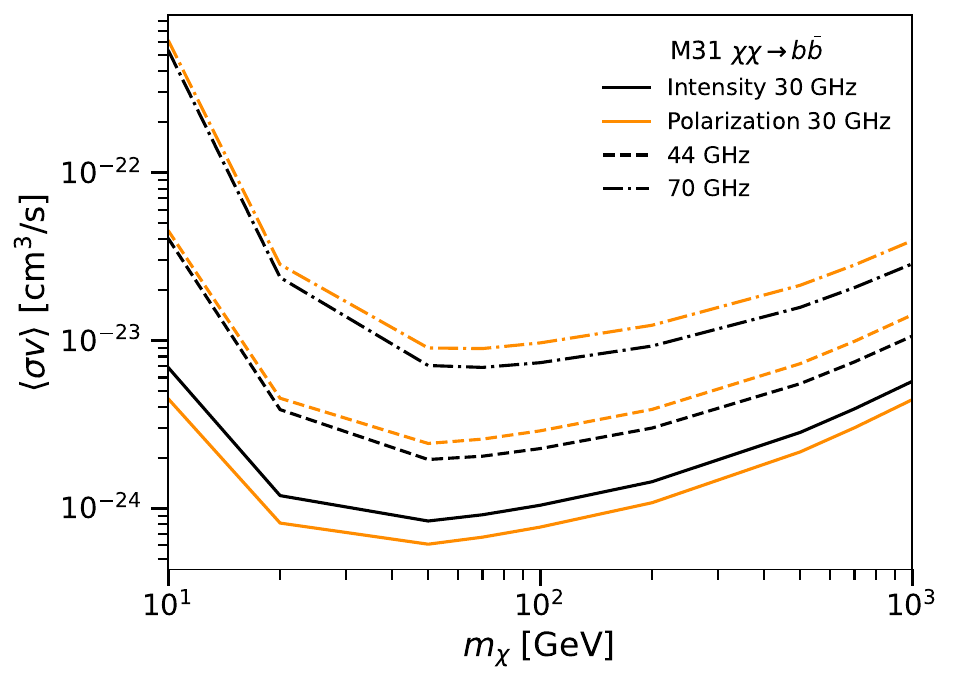}
  \hfill
  \includegraphics[width=0.32\textwidth]{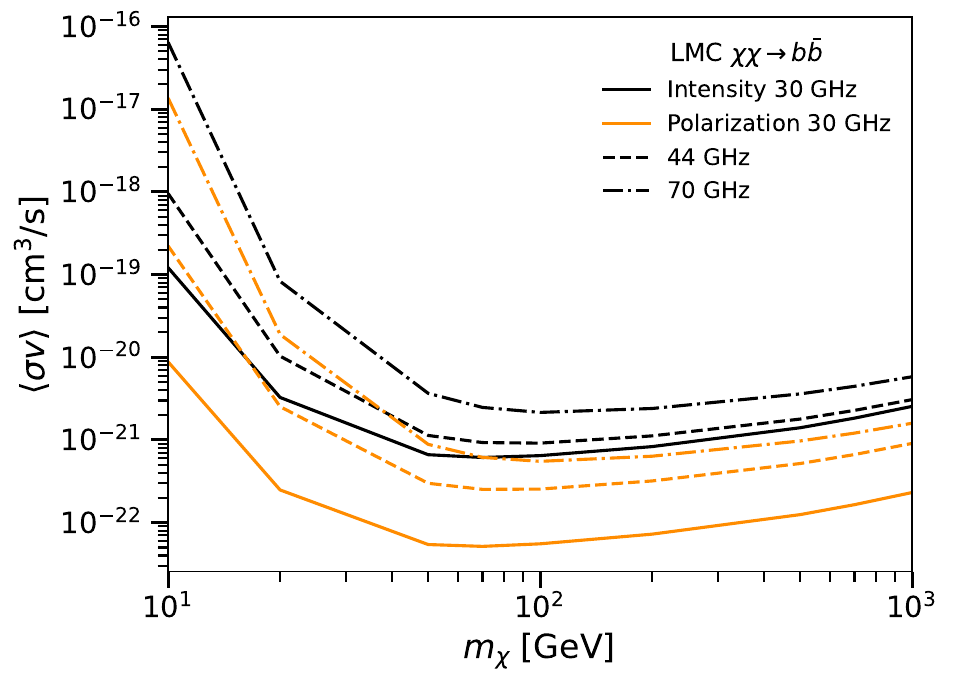}
  \hfill
  \includegraphics[width=0.32\textwidth]{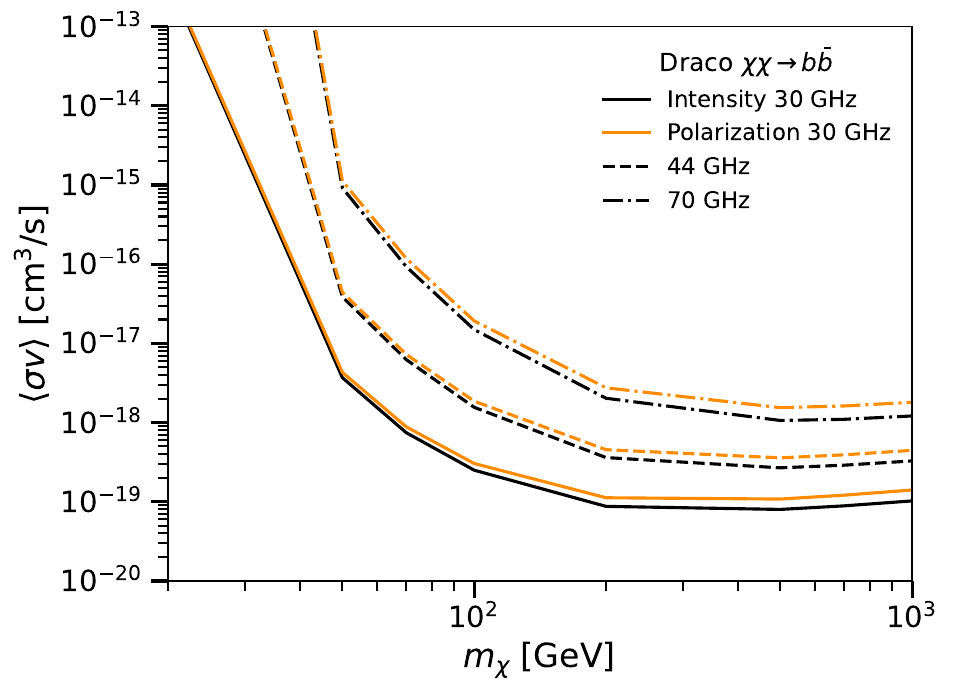}
  \\[1em]
  \hspace*{0.16\textwidth}
  \includegraphics[width=0.32\textwidth]{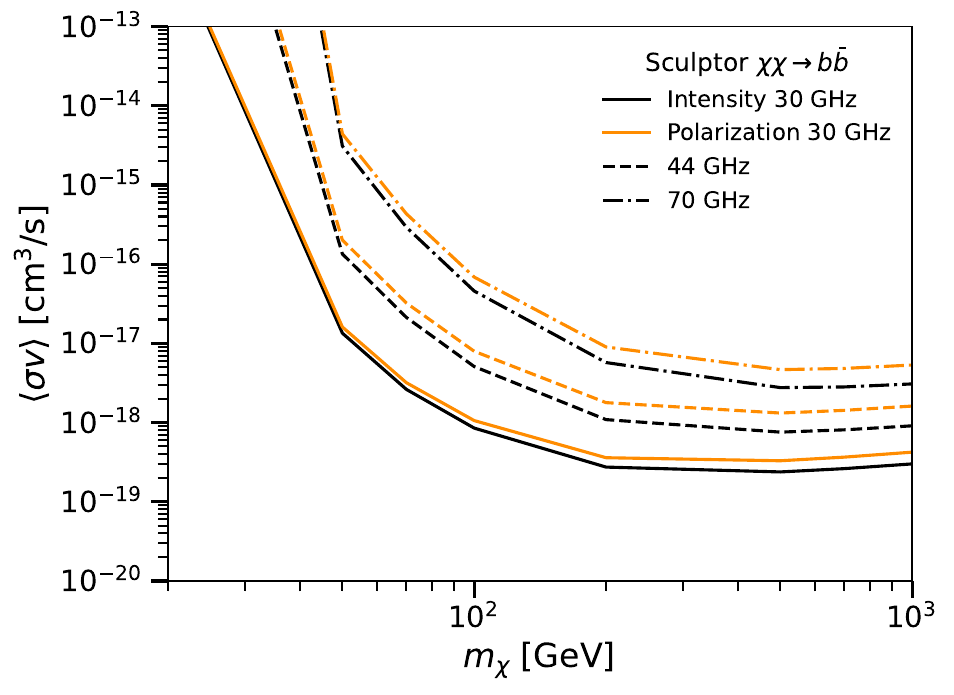}
  \hfill
  \includegraphics[width=0.32\textwidth]{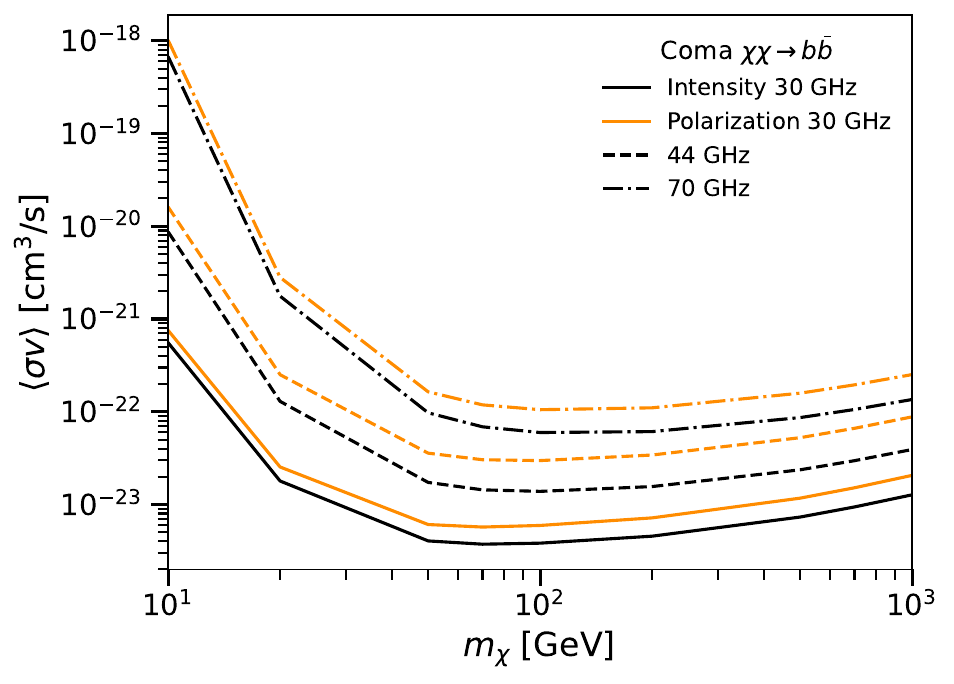}
  \hspace*{0.16\textwidth}
  \caption{$95\%$~CL upper limits on the DM annihilation cross section
    $\langle\sigma v\rangle$ as a function of DM mass $m_\chi$ for
    annihilation into $b\bar{b}$, for all five targets: M31, LMC,
    Draco, Sculptor, and Coma (left to right, top to bottom).  Solid,
    dashed, and dash-dotted lines show 30, 44, and 70~GHz results
    respectively; black lines correspond to total-intensity limits and
    orange lines to polarized-intensity limits.}
  \label{fig:limits_annihilation_bb}
\end{figure}

\begin{figure}[t]
  \centering
  \includegraphics[width=0.32\textwidth]{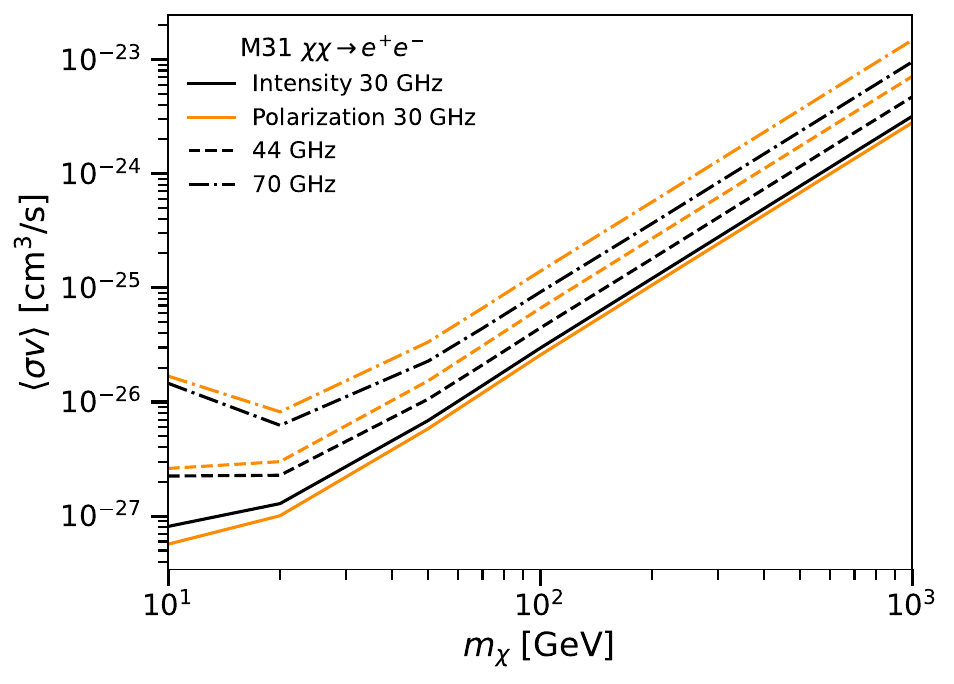}
  \hfill
  \includegraphics[width=0.32\textwidth]{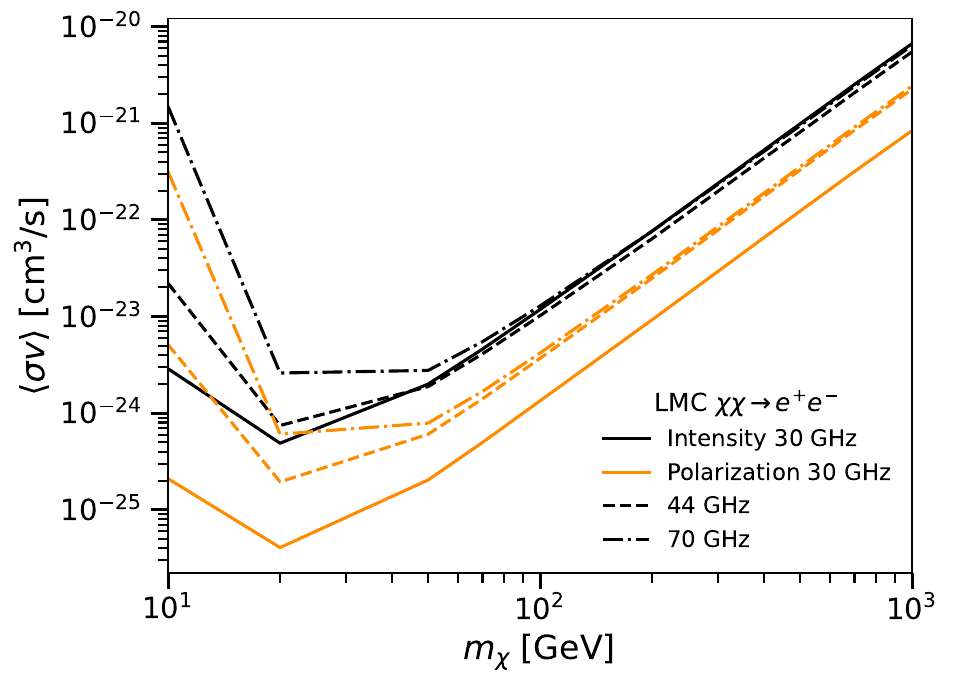}
  \hfill
  \includegraphics[width=0.32\textwidth]{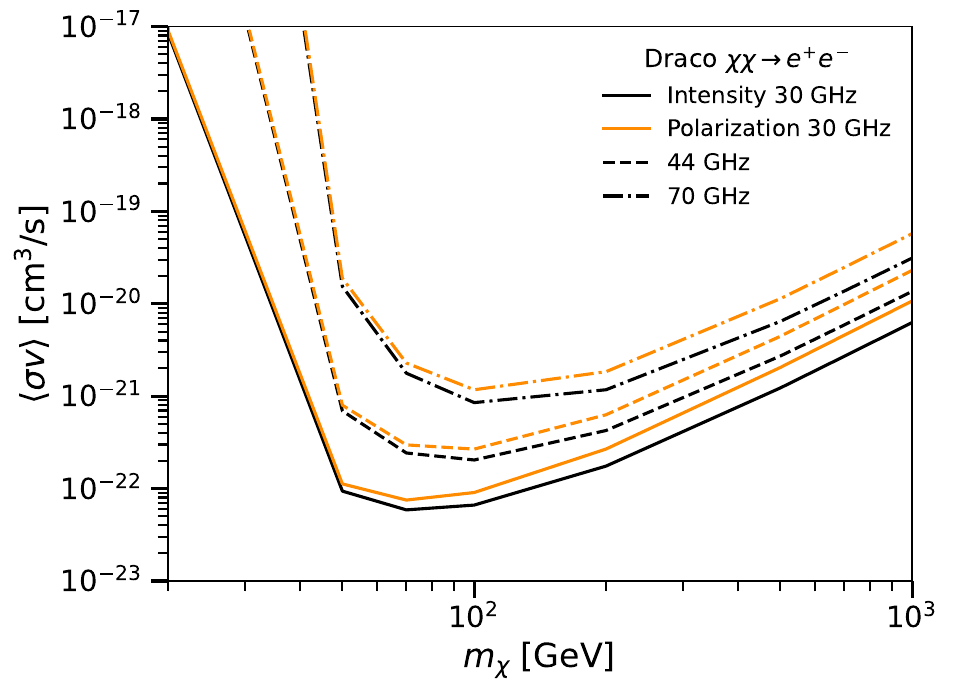}
  \\[1em]
  \hspace*{0.16\textwidth}
  \includegraphics[width=0.32\textwidth]{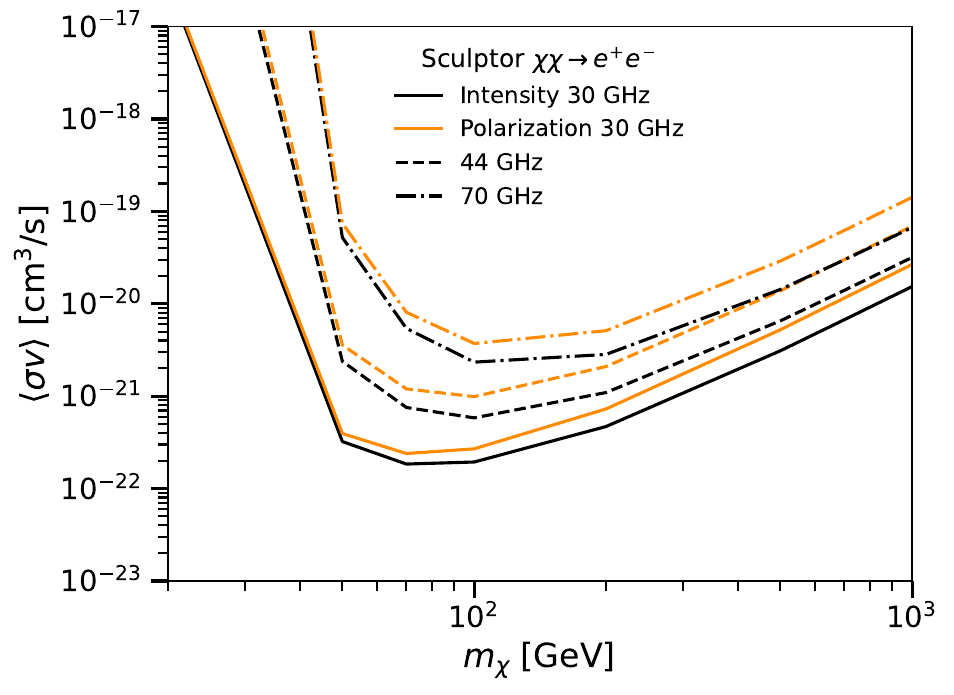}
  \hfill
  \includegraphics[width=0.32\textwidth]{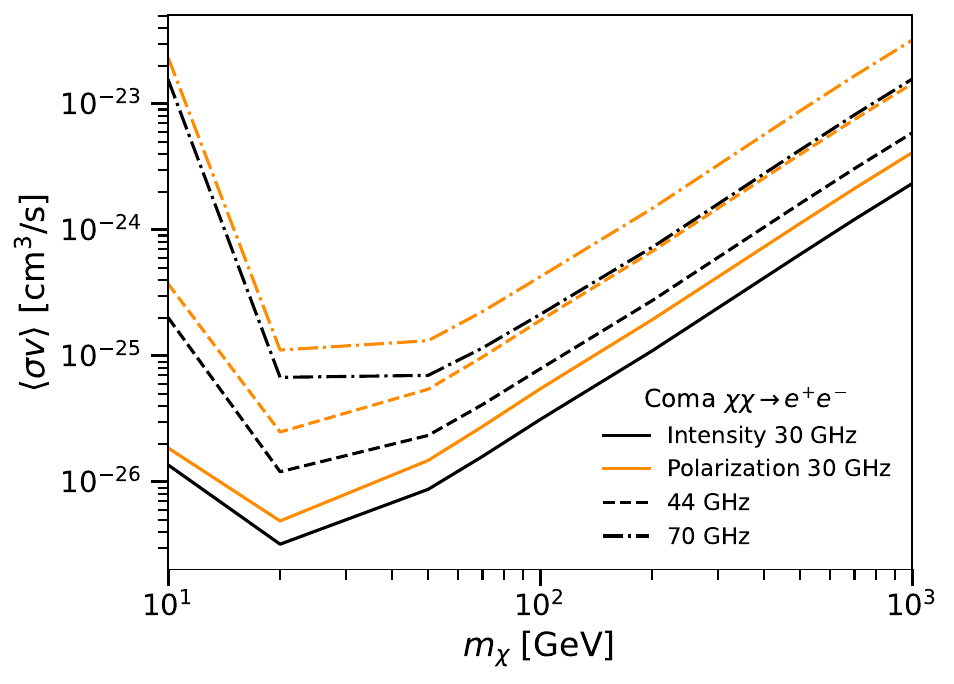}
  \hspace*{0.16\textwidth}
  \caption{Same as Figure~\ref{fig:limits_annihilation_bb} but for
    annihilation into $e^+e^-$.  Limits are uniformly stronger than
    for $b\bar{b}$ because the harder leptonic spectrum produces more
    synchrotron power at the Planck frequencies.}
  \label{fig:limits_annihilation_ee}
\end{figure}

\begin{figure}[t]
  \centering
  \includegraphics[width=0.32\textwidth]{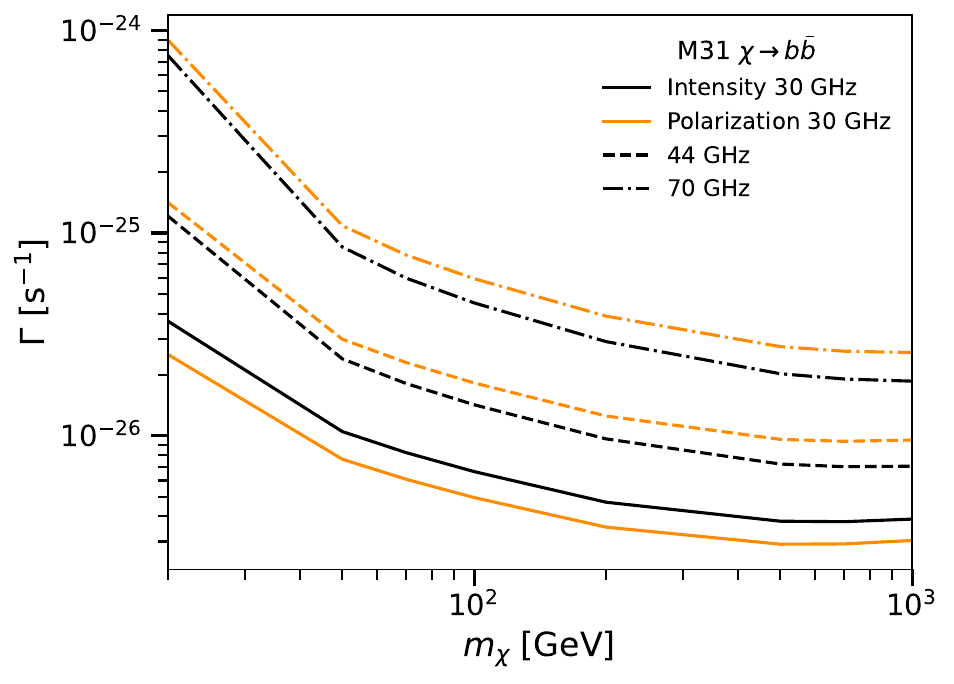}
  \hfill
  \includegraphics[width=0.32\textwidth]{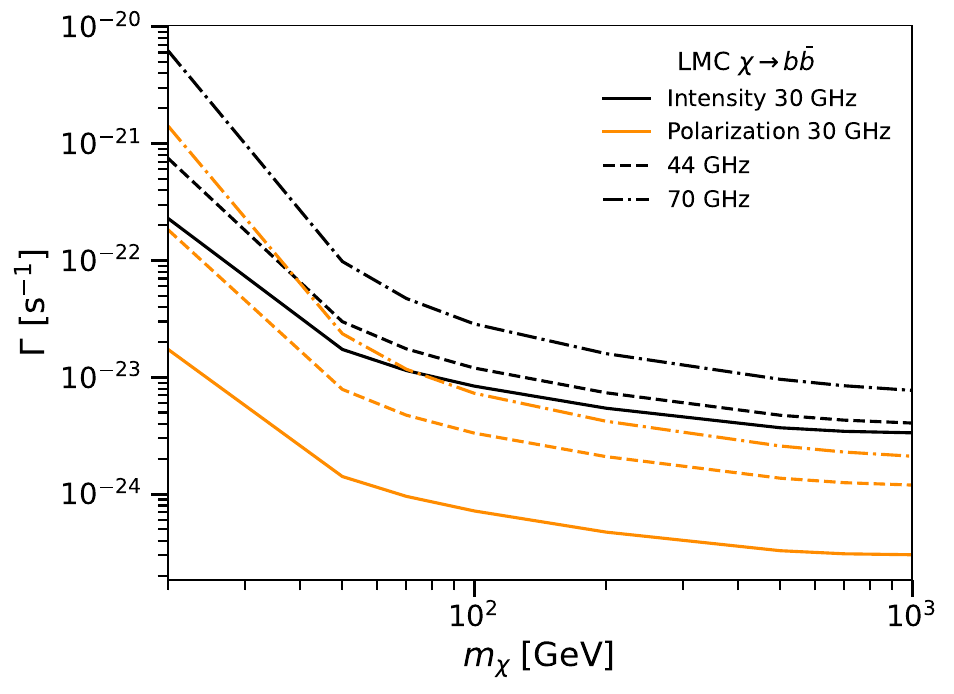}
  \hfill
  \includegraphics[width=0.32\textwidth]{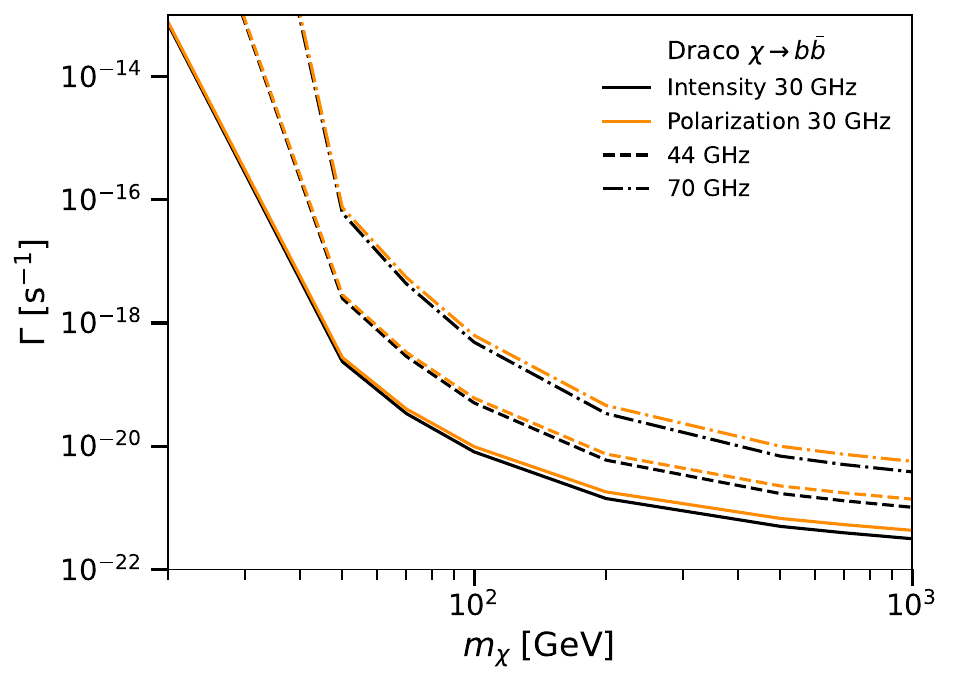}
  \\[1em]
  \hspace*{0.16\textwidth}
  \includegraphics[width=0.32\textwidth]{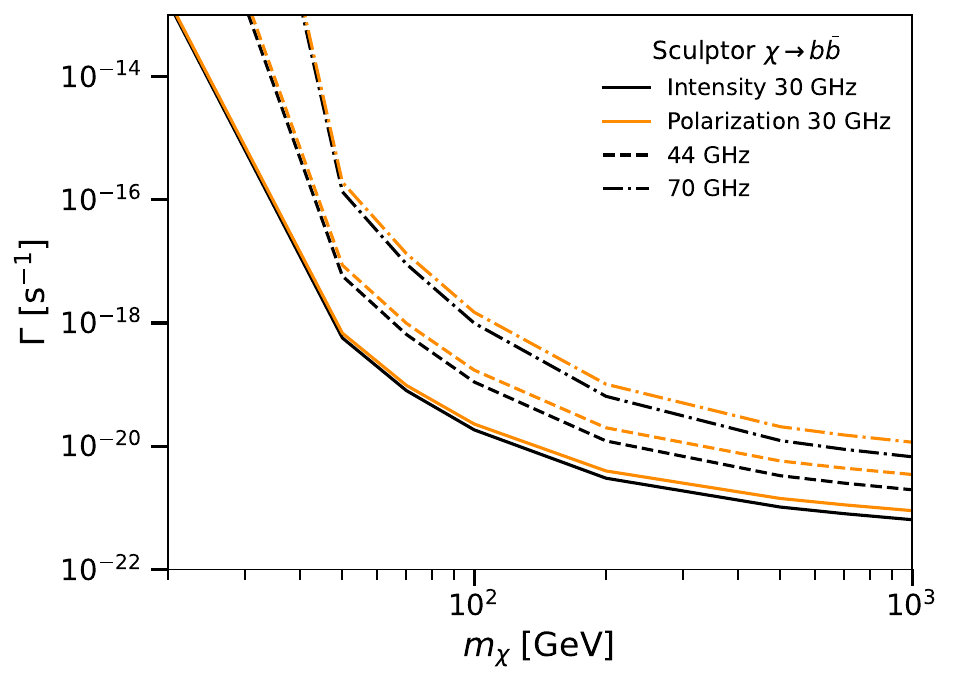}
  \hfill
  \includegraphics[width=0.32\textwidth]{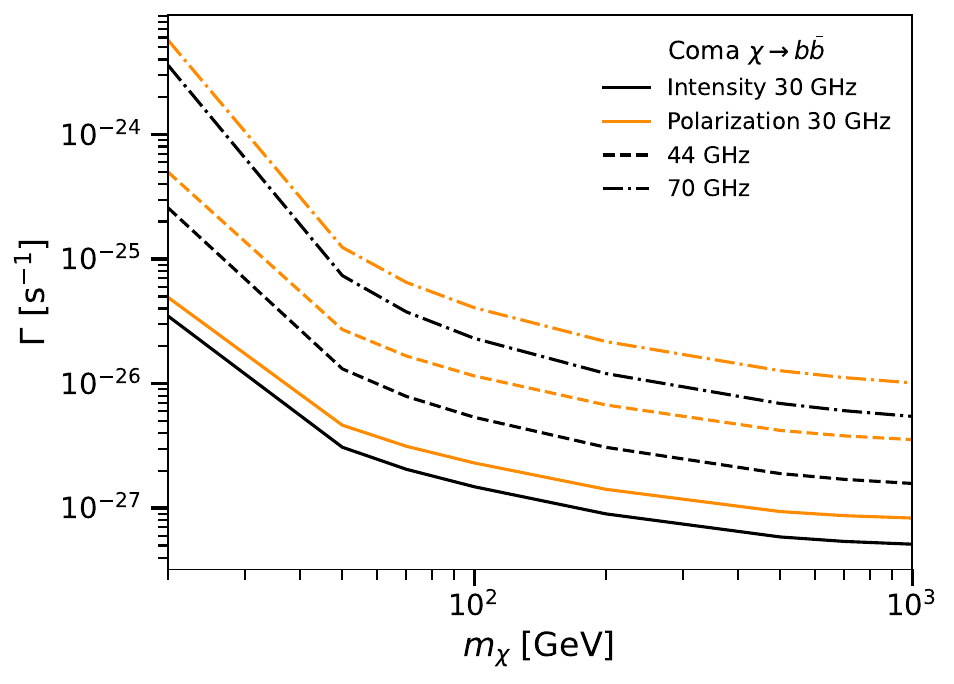}
  \hspace*{0.16\textwidth}
  \caption{$95\%$~CL lower limits on the DM lifetime $\tau = 1/\Gamma$
    as a function of $m_\chi$ for decay into $b\bar{b}$, for all five
    targets: M31, LMC, Draco, Sculptor, and Coma (left to right, top
    to bottom).  Line styles and colors are as in
    Figure~\ref{fig:limits_annihilation_bb}.}
  \label{fig:limits_decay_bb}
\end{figure}

\begin{figure}[t]
  \centering
  \includegraphics[width=0.32\textwidth]{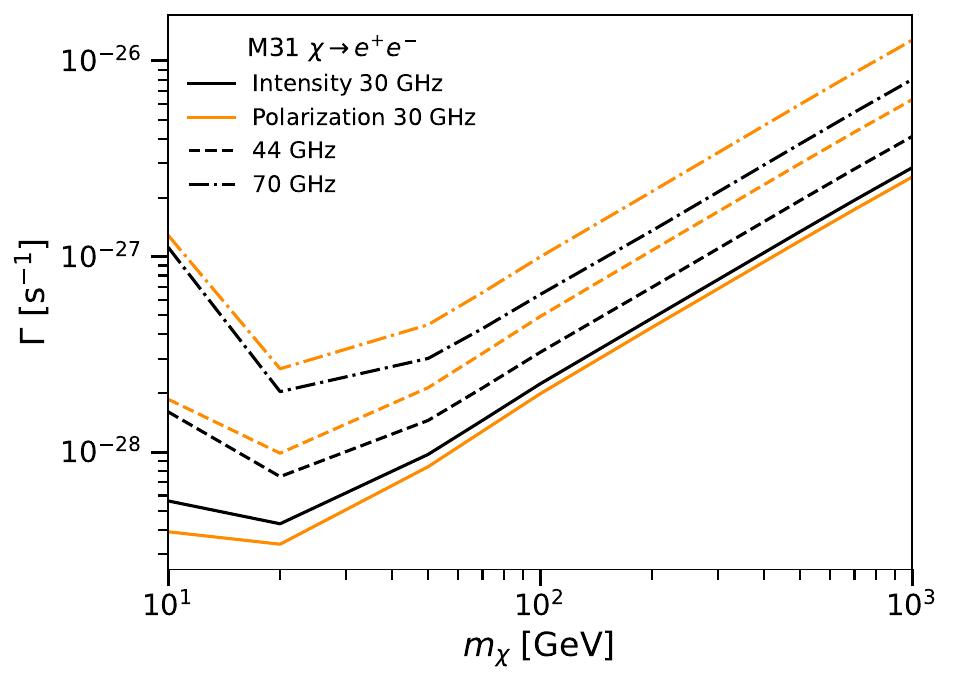}
  \hfill
  \includegraphics[width=0.32\textwidth]{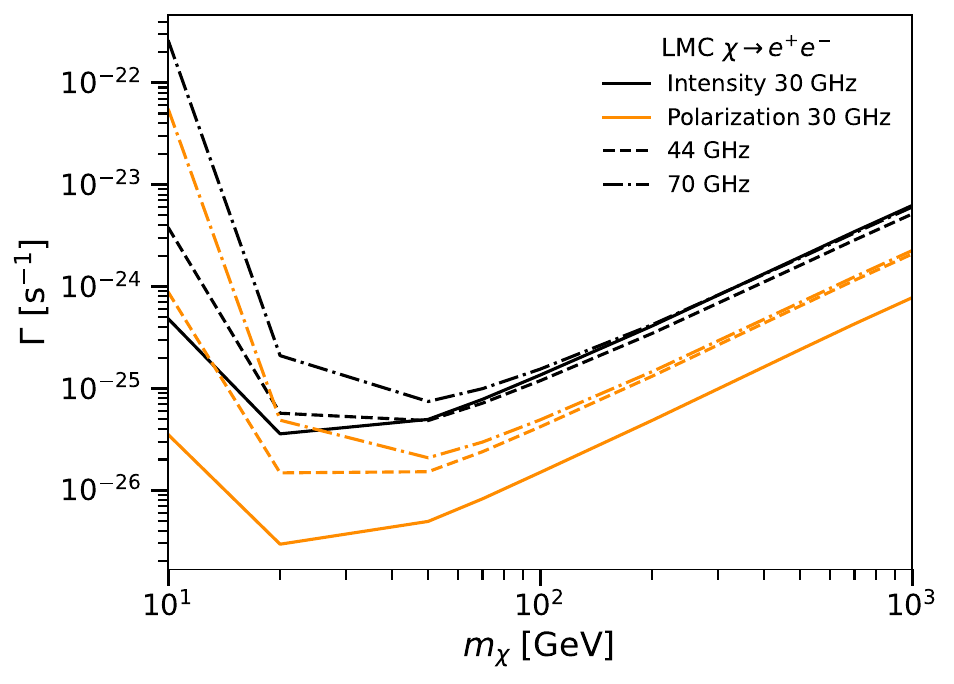}
  \hfill
  \includegraphics[width=0.32\textwidth]{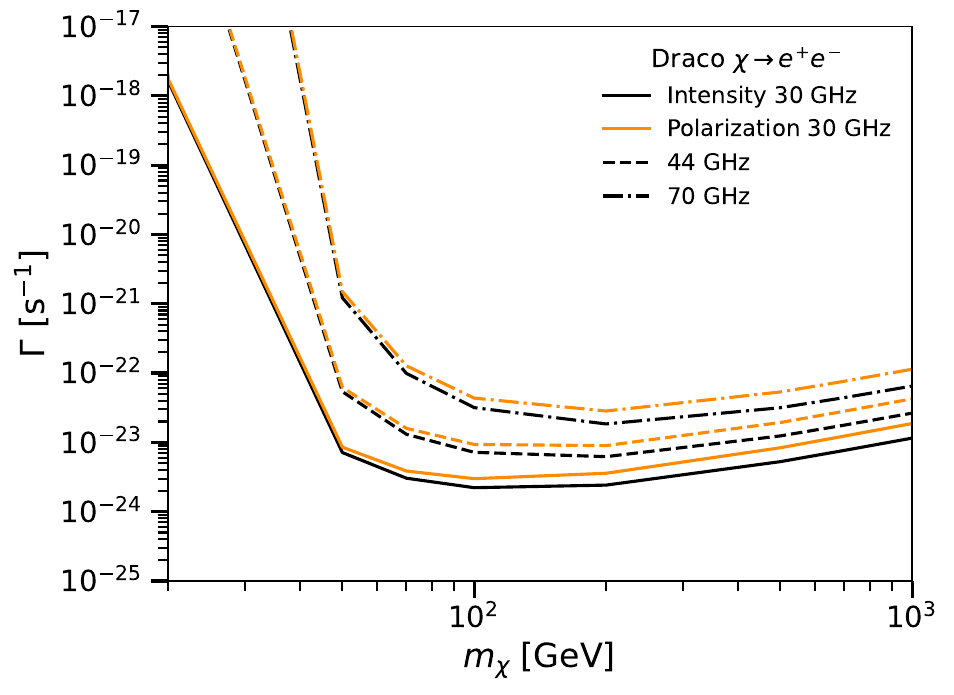}
  \\[1em]
  \hspace*{0.16\textwidth}
  \includegraphics[width=0.32\textwidth]{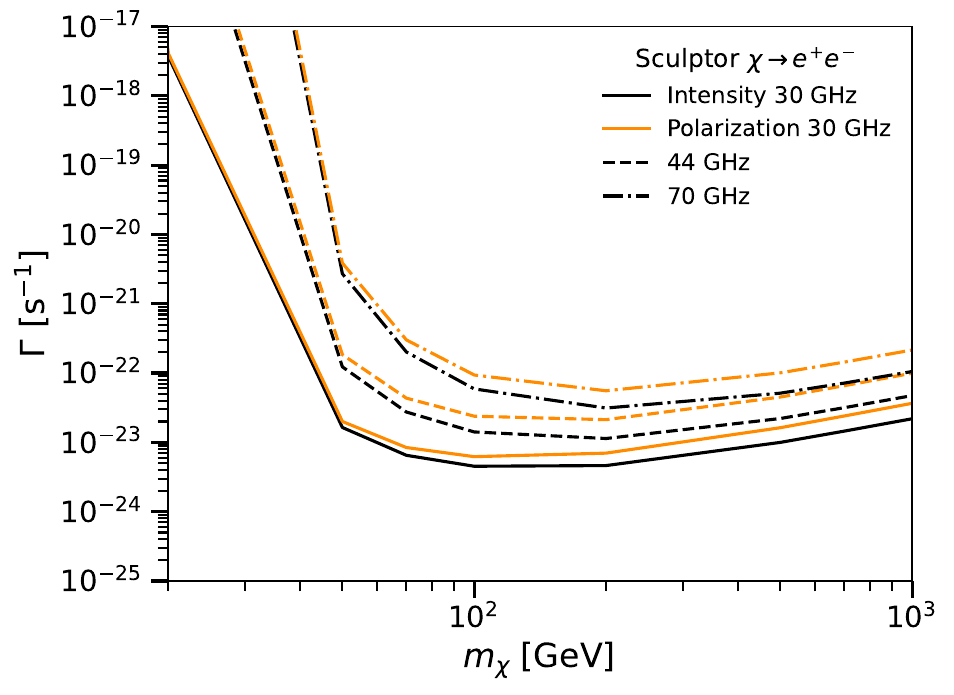}
  \hfill
  \includegraphics[width=0.32\textwidth]{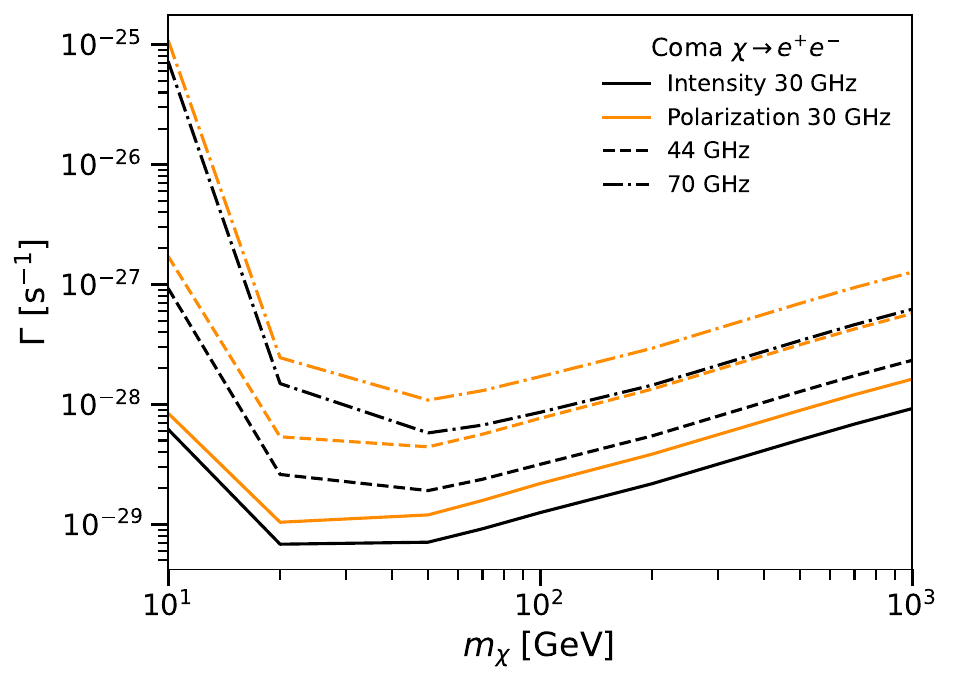}
  \hspace*{0.16\textwidth}
  \caption{Same as Figure~\ref{fig:limits_decay_bb} but for decay into
    $e^+e^-$.  As for annihilation, the leptonic channel yields
    stronger constraints than the hadronic one across all targets and
    frequencies.}
  \label{fig:limits_decay_ee}
\end{figure}

The $95\%$~CL upper limits are shown in
Figures~\ref{fig:limits_annihilation_bb}--\ref{fig:limits_decay_ee}.
In all figures, solid, dashed, and dash-dotted lines correspond to 30,
44, and 70~GHz respectively; black lines show total-intensity limits
and orange lines show polarized-intensity limits.
 
\textit{Annihilation (Figs.~\ref{fig:limits_annihilation_bb}
and~\ref{fig:limits_annihilation_ee}).}
Figures~\ref{fig:limits_annihilation_bb}
and~\ref{fig:limits_annihilation_ee} show the limits on
$\langle\sigma v\rangle$ as a function of $m_\chi$ for the $b\bar{b}$
and $e^+e^-$ channels respectively.  Across all five targets, the
30~GHz channel provides the most stringent constraints, a consequence
of the higher per-pixel sensitivity of the Planck LFI maps at this
frequency for diffuse emission.  The $e^+e^-$ channel is constrained
more tightly than $b\bar{b}$ at every target: the harder $e^+e^-$
injection spectrum produces more synchrotron power at the Planck
frequencies for a given DM mass and cross section.  Among the five
targets, the dSphs (Draco and Sculptor) yield the most competitive
limits per unit $J$-factor owing to their negligible astrophysical
backgrounds, while the LMC and M31 benefit from large $J$-factors
despite higher foreground emission.  Coma provides limits broadly
comparable to the dSphs at high masses where the calorimetric regime
applies.
 
\textit{Decay (Figs.~\ref{fig:limits_decay_bb}
and~\ref{fig:limits_decay_ee}).}
Figures~\ref{fig:limits_decay_bb} and~\ref{fig:limits_decay_ee} show
the corresponding lower limits on the DM lifetime $\tau = 1/\Gamma$
for decay into $b\bar{b}$ and $e^+e^-$ respectively.  The trends
mirror the annihilation case: 30~GHz dominates, $e^+e^-$ is more
constraining than $b\bar{b}$, and the frequency ordering of the
three channels is preserved across all targets and mass ranges.  The
lifetime limits extend to $\tau \gtrsim 10^{25}$--$10^{28}$~s
depending on target and channel, competitive with gamma-ray and CMB
constraints at sub-TeV masses.
 
\textit{Total intensity versus polarization.}
For M31, Coma, Draco, and Sculptor, the total-intensity and
polarized-intensity limits are broadly comparable, with the
total-intensity limits marginally more stringent at most masses.
This is expected: the intrinsic polarization fraction is reduced by
field-disorder depolarization (Eq.~\ref{eq:Pi_disorder}), so $P <
\Pi_0 I$, but the Planck polarization noise is also lower than the
total-intensity noise in the diffuse-foreground-dominated regime,
partially compensating.
 
The LMC is the exception, as discussed in Section~\ref{sec:lmc_radio}.
Faraday depolarization in the turbulent LMC disk suppresses the
polarized signal at centimetre wavelengths, so the Planck polarized
maps are intrinsically fainter than the total-intensity maps at
30--70~GHz.  As a result, the upper limits derived from the polarized
maps are more stringent than those from total intensity for the LMC —
the polarized signal is quieter and therefore easier to beat with a
DM signal of given strength.  This makes total intensity the primary
estimator for the LMC limits, with the polarized maps providing the
most conservative bound.  Since DM annihilation injects $e^+/e^-$
pairs isotropically into the turbulent disk, any resulting synchrotron
signal is subject to the same depolarization as the disk's own
emission, further motivating total intensity as the primary probe for
this target.
 
\section{Conclusions}
\label{sec:conclusions}
 
We have presented the first systematic study of DM-induced synchrotron
emission in extragalactic targets using both total-intensity and
polarized Planck microwave maps simultaneously.  Our analysis covers
five targets spanning three orders of magnitude in halo mass — M31,
the Large Magellanic Cloud (LMC), the classical dSphs Draco and
Sculptor, and the Coma cluster — and derives 95\%~CL upper limits on
the DM annihilation cross section $\langle\sigma v\rangle$ and decay
rate $\Gamma$ as a function of DM mass $m_\chi$ for both $b\bar{b}$
and $e^+e^-$ final states.  We summarize the main results,
methodological advances, and astrophysical lessons below.
 
\paragraph{Methodology.}
The analysis is built on a self-consistent numerical pipeline in which
the steady-state $e^\pm$ phase-space density is solved with
\texttt{DRAGON} using target-specific magnetic-field, ISRF, and gas
models, and the resulting synchrotron emission is integrated along the
line of sight with \texttt{HERMES} to produce both Stokes~$I$ and
polarized-intensity $P = \sqrt{Q^2+U^2}$ maps.  The new
\texttt{HERMES} module for polarized synchrotron, implemented
following the \texttt{GALPROP}/\texttt{Hammurabi} scheme
\cite{Waelkens2009,Steininger2019}, allows the two Stokes observables
to be computed within a single framework for the first time for
extragalactic DM targets.  For M31 and the LMC, whose disk planes are
inclined relative to the line of sight, the astrophysical inputs are
rotated into the galactic frame before transport and then rotated back
for the Planck comparison, extending the procedure of
\citet{Reynoso-Cordova:2025meg} to polarization.  A Gaussian
single-bin likelihood is applied to the per-pixel Planck maps, with
three independent signal estimators (pixel value at nominal coordinate,
$0.5^\circ$-aperture average, and $0.5^\circ$-aperture maximum)
providing cross-checks of robustness; the three estimators agree to
within a factor of $\sim 2$ for all targets and frequencies.
 
\paragraph{Frequency and channel dependence.}
Across all five targets and both observables, the 30~GHz channel
provides the most stringent constraints, reflecting the higher
sensitivity of the Planck LFI maps at this frequency for diffuse
synchrotron emission.  The 44 and 70~GHz channels are weaker by
factors of a few to $\sim 10$ at low masses, with the gap narrowing
at high masses as the synchrotron peak shifts to higher frequencies.
The $e^+e^-$ channel yields limits uniformly stronger than $b\bar{b}$
by one to several orders of magnitude, because the harder leptonic
injection spectrum produces proportionally more synchrotron power at
Planck frequencies for the same DM mass and cross section.  This
hierarchy is a direct consequence of the emissivity scaling
$j_\nu \propto B_{\rm tot}^{(p+1)/2}\nu^{-(p-1)/2}$: a harder
electron spectrum ($p$ smaller) means more flux at a given frequency
per unit DM signal.
 
\paragraph{Total intensity versus polarization.}
For four of the five targets — M31, Coma, Draco, and Sculptor — the
total-intensity and polarized-intensity limits are broadly comparable,
differing typically by less than a factor of two at a given frequency
and mass.  This near-equality arises because two competing effects
roughly cancel: the polarized signal is intrinsically suppressed by
the field-disorder depolarization factor
$\Pi = \Pi_0 B_{\rm ord}^2 / B_{\rm tot}^2$
(Eq.~\ref{eq:Pi_disorder}), which is $\lesssim 1$, but the Planck
polarization noise in the diffuse-foreground-dominated regime is also
lower than the total-intensity noise, partially recovering the
sensitivity.  The result is that microwave polarimetry provides a
genuinely independent and complementary constraint channel: systematic
uncertainties in astrophysical foregrounds, which dominate
total-intensity maps, do not enter the polarization limits in the same
way, since the polarized foreground (dominated by Galactic synchrotron
at these frequencies) has a different spatial morphology and spectral
behavior from any DM signal.
 
\paragraph{The LMC as a special case.}
The LMC breaks the near-equality pattern in the opposite direction to
what one might expect.  The turbulence-dominated magnetic field of the
LMC disk ($\langle B \rangle \approx 10.1\,\mu$G, of which the
isotropic turbulent component dominates \cite{Hassani2022}) acts as an
efficient Faraday depolarizer at centimetre wavelengths
\cite{Gaensler2005}, biasing the observed polarized emission toward the
low-emissivity near-side halo and away from the star-forming disk
where the DM signal would be concentrated.  As a result, the Planck
polarized maps of the LMC are intrinsically quieter than the
total-intensity maps, and the polarized limits are more
\emph{stringent} — not weaker — than the total-intensity limits for
this target.  But the quieter polarized maps do not reflect a genuine
DM signal advantage: DM annihilation injects $e^+/e^-$ pairs
isotropically, producing synchrotron emission with no intrinsic ordered
component, which would be subject to the same Faraday depolarization.
The total-intensity limits are therefore the physically relevant ones
for the LMC, with the polarized maps providing the most conservative
bound.  This distinction — between astrophysical depolarization
suppressing the background vs.\ the same suppression affecting the
signal — is a general lesson for polarimetric DM searches in
turbulent, star-forming environments.
 
\paragraph{Comparison with prior work.}
The only previous use of synchrotron polarization for DM constraints
is the Galactic-halo analysis of Manconi, Cuoco \& Lesgourgues
\cite{Manconi2022}, who found that Planck polarization maps improve
on total-intensity limits by roughly an order of magnitude for
leptophilic DM.  That gain does not carry over to extragalactic
targets at Planck angular resolution: the beam ($\sim 13'$--$32'$)
subtends several kpc in all five targets, so beam depolarization
from unresolved turbulence is a dominant suppression mechanism that
was less severe in the Galactic analysis, which benefits from
kiloparsec-scale resolution toward high-latitude regions with ordered
fields.  For the four non-LMC extragalactic targets, total intensity
and polarization are thus comparable probes, whereas in the Galactic
case polarization wins decisively.  The extragalactic regime
therefore represents a qualitatively different operating point of
polarimetric DM searches, and the two approaches are complementary
rather than redundant.
 
For total intensity, our limits at 30~GHz for the dSphs are
competitive with dedicated GHz-band radio observations
\cite{Regis2015II,Regis2015III} at similar DM masses, despite the
much lower angular resolution of Planck; the advantage of Planck is
the simultaneous access to polarization and the all-sky coverage that
enables the five-target joint analysis in a single pipeline.  The LMC
total-intensity limits from this work are weaker than the dedicated
ASKAP-EMU analysis \cite{Regis2021LMC} at lower frequencies and
higher angular resolution, which is expected: the synchrotron
emissivity is higher at 888~MHz, and the much smaller Planck beam
at 30~GHz ($32'$) dilutes the signal compared to the $14''$
ASKAP resolution.  Our contribution for the LMC is the first
polarimetric treatment of its DM signal and the first demonstration
that total intensity and polarization limits can be simultaneously
derived within a consistent framework.
 
\paragraph{Astrophysical systematics.}
The dominant sources of systematic uncertainty in these results are
the magnetic-field models and the DM density profiles.  For the dSphs,
the field is unconstrained by direct measurement and is assumed
uniform at $B = 0.3\,\mu$G; varying $B$ across $0.1$--$1\,\mu$G
shifts the limits by less than a factor of two, consistent with the
CMB-loss-dominated transport regime where $j_\nu \propto B^{(p+1)/2}$
but losses are $IC$-dominated.  For M31, the Beck \& Berkhuijsen
\cite{Beck2025} equipartition analysis anchors the field to
$B_{\rm tot} = 6.3 \pm 0.2\,\mu$G in the emission torus, giving
well-controlled synchrotron emissivity there.  The NFW profile
uncertainties for the dSphs are mitigated by using posterior median
values from dedicated kinematic analyses
\cite{Regis:2023rpm,Todarello:2023hdk} rather than generic
parameterizations.  The Planck noise covariance is the dominant
statistical uncertainty, and our results are robust at the factor-of-2
level against the specific flux estimator and aperture choice.
 
\paragraph{Outlook.}
This work demonstrates that polarimetric DM searches with microwave
all-sky surveys are feasible and competitive with dedicated radio
observations for select targets.  Several directions offer
significant improvement.  For dSphs, the sub-arcminute resolution
and $\mu$Jy sensitivity of MeerKAT and the ngVLA would resolve the
DM emission profile and separate it from point sources, delivering
limits orders of magnitude stronger than those possible with Planck.
For galaxy clusters, wide-band polarimetry with the SKA — where
rotation-measure synthesis can be applied — would probe the
frequency-dependent depolarization of DM-induced emission and
potentially distinguish it from astrophysical cosmic-ray backgrounds
through its distinct morphology (peaked at the halo center, not at
radio-bright substructures).  For the LMC and M31, high-resolution
radio polarimetry at $\gtrsim 5$~GHz, where internal Faraday
depolarization is reduced, would recover the polarized DM signal that
is suppressed at centimetre wavelengths, potentially enabling the
order-of-magnitude gain over total intensity seen by Manconi et al.\
\cite{Manconi2022} in the Galactic case.  More broadly, the
target-specific treatment of magneto-ionic environments developed here
— with self-consistent gas, ISRF, and magnetic-field models for each
system — provides a template for systematic polarimetric DM searches
across the diverse range of environments that future radio observatories
will access.

\begin{acknowledgments}
We thank Pedro de la Torre Luque for useful discussion. S.P. acknowledges support the U.S.\ Department of Energy grant number de-sc0010107.  JRC acknowledges the support of the Natural Sciences and Engineering Research Council of Canada (NSERC), funding reference number RGPIN-2020-07138, and the NSERC Discovery Launch Supplement, DGECR-2020-00231. JRC acknowledges support from INFN through the Senior Research Fellowship program (Grant No. 27076).
\end{acknowledgments}

\bibliographystyle{apsrev4-2}
\bibliography{dmsynchro}

\end{document}